\documentclass[twocolumn,
aps,
prx,
superscriptaddress,
showpacs,
amsmath,
amssymb,
floatfix,
longbibliography,
eprint]
{revtex4-2}
\usepackage{appendix}
\usepackage{xcolor}
\definecolor{darkblue}{RGB}{0,0,150}
\definecolor{nightblue}{RGB}{0,0,100}
\usepackage{graphicx,mathtools,bm,bbm}
\usepackage{MnSymbol}
\usepackage[
colorlinks,
citecolor=magenta,
linkcolor=darkblue,
urlcolor=nightblue]{hyperref}
\usepackage[english]{babel}
\usepackage[babel,kerning=true,spacing=true]{microtype}
\usepackage[utf8]{inputenc}
\usepackage{soul}

\DeclareMathOperator{\sgn}{sgn}
\DeclareMathOperator{\arctanh}{arctanh}
\renewcommand{\Im}{\ensuremath{\mathrm{Im}\,}}
\renewcommand{\Re}{\ensuremath{\mathrm{Re}\,}}
\newcommand{\refsub}[2]{\hyperref[#1]{\ref*{#1}#2}}
\newcommand{\mytitle}{Unconventional superconductivity in systems with annular Fermi surfaces: \strut\\\strut Application to rhombohedral trilayer graphene}

\begin{document}
    
\title{\mytitle}
    
\author{Areg Ghazaryan}
\thanks{A.G. and T.H. have contributed equally to this work.}
\affiliation{IST Austria, Am Campus 1, 3400 Klosterneuburg, Austria}
\author{Tobias Holder}
\thanks{A.G. and T.H. have contributed equally to this work.}
\affiliation{Department of Condensed Matter Physics, Weizmann Institute of Science, Rehovot 76100, Israel}
\author{Maksym Serbyn}
\affiliation{IST Austria, Am Campus 1, 3400 Klosterneuburg, Austria}
\author{Erez Berg}
\affiliation{Department of Condensed Matter Physics, Weizmann Institute of Science, Rehovot 76100, Israel}

\begin{abstract}
We show that in a two-dimensional electron gas with an annular Fermi surface, long-range Coulomb interactions can lead to unconventional superconductivity by the Kohn-Luttinger mechanism. Superconductivity is strongly enhanced when the inner and outer Fermi surfaces are close to each other. The most prevalent state has chiral p--wave symmetry, but d--wave and extended s--wave pairing are also possible. We discuss these results in the context of rhombohedral trilayer graphene, where superconductivity was recently discovered in  regimes where the normal state has an annular Fermi surface. Using realistic parameters, our mechanism can account for the order of magnitude of $T_c$, as well as its trends as a function of electron density and perpendicular displacement field. 
Moreover, it naturally explains some of the outstanding puzzles in this material, that include the weak temperature dependence of the resistivity above $T_c$, and the proximity of spin singlet superconductivity to the ferromagnetic phase.
\end{abstract}

\maketitle

\emph{Introduction.}---Graphene based two-dimensional materials offer
a unique platform to study correlated electron phenomena with an unprecedented
level of control~\cite{feldman2009broken,mayorov2011interaction,cao2018correlated,cao2018unconventional,lu2019superconductors,liu2019spin,Yankowitz1059,chen2019signatures,hao2021electric,park2021tunable}. Among these, rhombohedral trilayer graphene
(RTG)~\cite{Koshino2009,MacDonald2010,Koshino2010,Mak2010} has recently been shown to exhibit a rich phase diagram, including
several spin and valley polarized states, tunable by varying
the electron density and perpendicular electric field~\cite{Zhou2021half}. Most
strikingly,  two distinct superconducting phases were discovered
in different regions of the phase diagram \cite{Zhou2021superconductivity}. The emergence of
superconductivity in this ultra-pure and tunable system calls for
a theoretical understanding, potentially shedding light on long-standing
problems in condensed matter physics and opening the way to future
applications. 

In this work, we examine the key experimental facts about superconductivity
in RTG, and identify two puzzles that seem difficult to reconcile with
conventional, phonon-mediated superconductivity. We argue that these
puzzles can be resolved if one assumes an unconventional mechanism
for superconductivity \cite{note-unconventional}. Interestingly, superconductivity is found in regimes where the normal-state Fermi surface (FS) has an annular shape, with an inner electron pocket and an outer hole pocket. We show that an annular FS is beneficial for an electronic mechanism for superconductivity \cite{KohnLuttinger,kagan2015anomalous,Raghu2010,maiti2013superconductivity,comment_two_band,Raghu2011,Chubukov2017}, driven by repulsive Coulomb interactions. 
Using a microscopic model for RTG with realistic parameters, we find that 
the most likely candidate for the superconducting order parameter is a chiral p-wave, 
followed by extended s-wave \cite{Mazin2008}. 

\emph{Superconductivity in RTG}.---
The band structure of RTG~\cite{Koshino2009,MacDonald2010} exhibits strong peaks in the density of states, tunable by a perpendicular electric field, at finite density of electron or holes. 
Therefore, 
upon doping RTG undergoes a cascade of phase transitions between different
spin and valley polarized phases~\cite{Zhou2021half}, reminiscent of
the observations in twisted bilayer
graphene \cite{Zondiner2019,Wong2020cascade}, despite the absence of a moir\'e lattice in RTG. 

Recently, two superconducting phases were discovered in the hole-doped
side, labelled by SC1 and SC2 \cite{Zhou2021superconductivity}. SC1 is found within the flavor-symmetric
phase. 
Its maximum critical
temperature is $T_{c1}\approx100\,{\rm mK}$. SC2 occurs within a fully spin-polarized,
valley-unpolarized half-metal, 
with a $T_{c2}\lesssim50\,{\rm mK}$.
In both phases, the coherence length is shorter than the mean free path, placing them in the clean limit. 
The in-plane critical magnetic field of SC1 is consistent
with the Pauli limit~\cite{Clogston,chandrasekhar1962note}, while for SC2 the critical field exceeds the limit by an order of magnitude. 

\begin{figure*}[t]
    \centering
    \includegraphics[width=1.99\columnwidth]{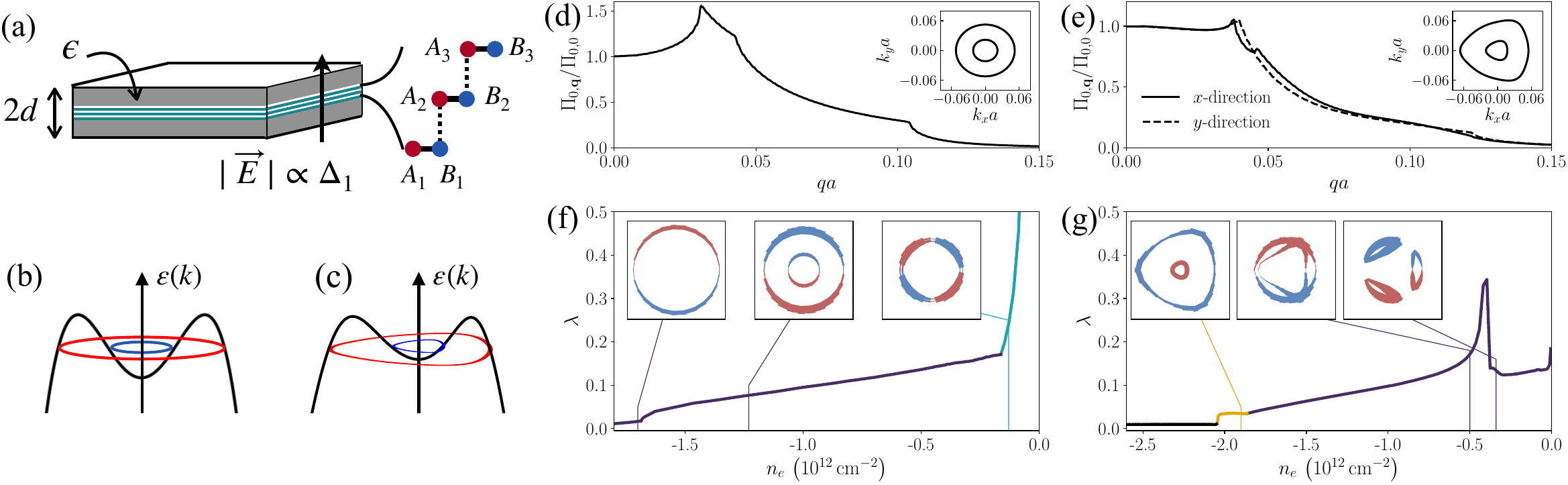}
    \caption{
    (a) RTG placed between two metallic gates at distance $d$ away, and encapsulated by an insulator with dielectric constant~$\epsilon$. (b,c) Band structure and FS of the simplified circularly symmetric model and the realistic model, respectively. (d,e) The polarization function $\Pi_{0,\bm{q}}$ within the simplified and realistic models, at density $n_e=-1.19\times10^{12}\,\mathrm{cm^{-2}}$ and $n_e=-1.67\times10^{12}\,\mathrm{cm^{-2}}$, respectively. (f,g) SC dimensionless coupling constant, $\lambda$, vs.~density in the two models, colored for p-wave (purple), extended s-wave (yellow) and d-wave (cyan). The inset shows the solution of the linearized gap equation along the FS at certain values of $n_e$. Red (blue) color represents positive (negative) $\Delta_{\bm{k}}$. We used 
    $\epsilon=4$ and $d=36.9\,\mathrm{nm}$ in (f-g) and $\Delta_1=20\,\rm meV$ in (e,g).
    }
    \label{fig1}
\end{figure*}

It has been proposed \cite{Chou2021} that RTG is a
`conventional' phonon-driven superconductor, with a s-wave like gap wave function 
within each valley. 
We point out two difficulties with this scenario. 
First, coupling to acoustic phonon modes should lead to a linear dependence of the resistivity for $T\gtrsim\Theta_{{\rm BG}}/4$, where $\Theta_{{\rm BG}}\approx40\,{\rm K}$ is the Bloch-Grüneisen temperature at the density range of SC1.
However, the resistance
of RTG above SC1 is nearly temperature-independent
up to $T=20\,{\rm K}$~\cite{Zhou2021superconductivity}. 
This is unlike many conventional superconductors, where estimates for the dimensionless electron-phonon coupling from $T_{c}$ and from the slope of the resistivity ${\rm vs.}~T$ typically agree~\cite{allen2000electron}. 
Second, SC1 emerges out of a spin- and valley-unpolarized normal state~\cite{Zhou2021half,Zhou2021superconductivity}. 
As explained below, SC1 can be either spin singlet or triplet, and one expects the inter-valley exchange coupling $J_{H}$ to determine which one is realized. 
Experimentally, SC1 is spin singlet, which implies $J_{H}<0$ for an s-wave like gap function.
This is in apparent contradiction with the presence of a spin polarized ferromagnetic
state nearby in the phase diagram, that requires $J_{H}>0$~\cite{Note1}.
We argue that both of these puzzles are resolved for an all-electronic mechanism that leads to unconventional superconductivity.

\emph{Model.}---
The Hamiltonian used to describe rhombohedral trilayer graphene reads
\begin{equation}
\hat{H}=\hat{H}_{0}+\hat{H}_{C},\label{eq:H}
\end{equation}
where 
$
\hat{H}_{0}=\sum_{\bm{k},\tau,s}\Psi_{\bm{k}\tau s}^{\dagger}h_{\bm{k}\tau}^{\vphantom{\dagger}}\Psi_{\bm{k}\tau s}^{\vphantom{\dagger}}
$
is the single-particle part~\cite{MacDonald2010}. $h_{\bm{k}\tau}$
is a $6\times6$ matrix in the basis $\left(A_1,B_1,A_2,B_2,A_3,B_3\right)$,
corresponding to the $A,B$ sublattices of layers 1,2,3 [Fig.~\refsub{fig1}{(a)}] that is written explicitly in the Supplemental Material~\cite{SI}.
The valley and spin indices are denoted by $\tau$ and $s$, respectively. $\Psi_{\bm{k}\tau s}^{\dagger}$
is a spinor containing the operators $\psi_{\bm{k},\ell,\sigma,\tau,s}^{\dagger}$,
that create an electron with momentum $\bm{k}$, at layer
$\ell=1,2,3$ and sublattice $\sigma=A,B$. The Coulomb interaction
between Fourier component of the density $\rho_{\bm{q}}=\sum_{\bm{k}\tau s}\Psi_{\bm{k}\tau s}^{\dagger}\Psi_{\bm{k}+\bm{q}\tau s}^{\vphantom{\dagger}}$ is given by
\begin{equation}
\hat{H}_{C}=\frac{1}{2L^{2}}\sum_{\bm{q}}V_{0,\bm{q}}\rho_{\bm{q}}\rho_{-\bm{q}},\label{eq:rhoq}
\end{equation}
where $V_{0,\bm{q}}=\frac{2\pi e^{2}}{\epsilon q}\tanh(qd)$ is the
Coulomb potential screened by two metallic gates at distance $d$
above and below the RTG, $\epsilon$ is the dielectric
constant, and $L^2$ is the area
of the system. 

Near the Fermi level, the band structure of $h_{\bm{k}\tau}$ consists of a conduction and a valence band separated by a gap $2\Delta_{1}$, proportional to the perpendicular displacement field. 
$\Delta_{1}$ and the carrier density $n_e$, controlled experimentally by gate voltages, tune the system between different phases.

We project the Hamiltonian onto the valence band, where SC1 and SC2 occur. 
This amounts to substituting $\rho_{\bm{q}}$ 
by $\tilde{\rho}_{\bm{q}}=\sum_{\bm{k},\tau,s}\Lambda^{\vphantom{\dagger}}_{\bm{k},\bm{q},\tau}c_{\bm{k},\tau,s}^{\dagger}c_{\bm{k}+\bm{q},\tau,s}^{\vphantom{\dagger}}$,
where $c_{\bm{k},\tau,s}^{\dagger}$ creates an electron at momentum $\bm{k}$
in the valence band in valley $\tau$ and spin $s$,
and $\Lambda_{\bm{k},\bm{q},\tau}=\langle u_{\bm{k},\tau}\vert u_{\bm{k}+\bm{q},\tau}\rangle$
is an overlap between Bloch wave functions of the valence
band, $\vert u_{\bm{k},\tau}\rangle$.

We investigate superconductivity within a purely electronic mechanism,
driven by the Coulomb interaction. Within this mechanism~\cite{KohnLuttinger,Raghu2010}, pairing is mediated by particle-hole
 fluctuations at a broad range of energies. The effective interaction after screening by
such fluctuations depends weakly on frequency, and will be treated
as instantaneous. Within the random phase approximation (RPA), this
interaction is given by
\begin{equation}
V_{\bm{q}}=\frac{V_{0,\bm{q}}}{1+\Pi_{0,\bm{q}}V_{0,\bm{q}}},\label{eq:RPA}
\end{equation}
where $\Pi_{0,\bm{q}}=N\sum_{\bm{k}}\left|\Lambda_{\bm{k},\bm{q},\tau}\right|^{2}\frac{f(\varepsilon_{\bm{k},\tau})-f(\varepsilon_{\bm{k}+\bm{q},\tau})}{\varepsilon_{\bm{k}+\bm{q},\tau}-\varepsilon_{\bm{k},\tau}}$
is the static polarization function, with $\varepsilon_{\bm{k},\tau}$ being the dispersion
of the valence band in valley $\tau$. $f(\varepsilon)$ is the Fermi function. $N=4$ is the number of spin and valley flavors. Note that $\Pi_{0,\bm{q}}$ is independent of $\tau$ due to time reversal symmetry. 

The RPA interaction Hamiltonian is given by $\hat H_{{\rm RPA}}=\frac{1}{2L^{2}}\sum_{\bm{q}}V_{\bm{q}}\tilde{\rho}_{\bm{q}}\tilde{\rho}_{-\bm{q}}$.
The superconducting $T_c$ is found by solving the
linearized BCS gap equation using
Eq.~(\ref{eq:RPA}) as the pairing interaction~\cite{SI}.
The gap equation reads
\begin{equation}
\Delta_{\bm{k}}=-\log\left(\frac{W}{T_{c}}\right)\int_{\rm{FS}}\frac{dk'_{\parallel}}{(2\pi)^{2}v_{\bm{k}'}}V_{\bm{k}-\bm{k}'} |\Lambda_{\bm{k},\bm{k}'-\bm{k},+1}|^2\Delta_{\bm{k}'},\label{eq:lin}
\end{equation}
where $\Delta_{\bm{k}}$ is the SC order parameter at a point $\bm{k}$
on the FS in valley $\tau=+1$, $v_{\bm{k}}$ is the magnitude of the Fermi
velocity at that point, and $W$ is an upper cutoff of the order of
the Fermi energy. The integration is taken over all FSs.
The largest eigenvalue $\lambda$ of the linear operator
$\mathcal{M}$ defined by $\mathcal{M}\Delta_{\bm{k}}=-\int\frac{dk'_{\parallel}}{(2\pi)^{2}v_{\bm{k}'}}V_{\bm{k}-\bm{k}'} |\Lambda_{\bm{k},\bm{k}'-\bm{k},+1}|^2 \Delta_{\bm{k}'}$,
yields $T_{c}=We^{-1/\lambda}$.

\emph{Kohn-Luttinger mechanism: idealized model.---}Since $V_{\bm{q}}>0$
for all $\bm{q}$, a solution $\Delta_{\bm{k}}$ of Eq.~(\ref{eq:lin})
 must change the sign across the FS. Therefore, 
the solution typically has a non s-wave symmetry, depending on the
structure of~$\Pi_{0,\bm{q}}$. For a single parabolic band in two
dimensions, $\Pi_{0,\bm{q}}$ is constant up to $q=2k_{F}$, and Eq.~(\ref{eq:lin}) has no solution~\cite{Chubukov1993}. In the case of two subbands with different Fermi momenta, the total polarization function is no longer a constant for momenta smaller than $2k_F$ of the outer FS, yielding non-trivial solutions~\cite{Raghu2011,Chubukov2017}.
In RTG, both SC1 and SC2 occur in regions where the normal state exhibits an annular FS, where the inner and outer Fermi surfaces have opposite Fermi velocities. As we demonstrate, such a dispersion is favorable for superconductivity. 

We begin with a simplified model where the dispersion is approximated as $\varepsilon_{\bm{k}}=-\varepsilon_{0}(k^{2}/k_{0}^{2}-1)^{2}-\mu$,
characterized by an energy scale $\varepsilon_{0}$ and a momentum
scale $k_{0}$, 
[Fig.~\refsub{fig1}{(b)}]. 
Moreover, we approximate $\Lambda_{\bm{k},\bm{q},\tau}=1$. 
Due to the rotational symmetry of this model, 
$\Pi_{0,\bm{q}}$ can be computed analytically~\cite{SI}. $\Pi_{0,\bm{q}}$ 
exhibits singularities for
$2k_{F1}$, $2k_{F2}$, and $k_{F1}-k_{F2}$, where $k_{F1,2}=k_{0}\sqrt{1\pm\sqrt{-\mu/\varepsilon_{0}}}$
are the momenta of the outer and inner FS, respectively, see Fig.~\refsub{fig1}{(d)}.

The result
for $\lambda$ as a function of electron density is shown in Fig.~\refsub{fig1}{(f)}. 
At large hole densities where the inner pocket disappears, $\lambda$ is nearly zero. $\lambda$ increases discontinuously once the inner pocket appears (around $n_e = -1.7\cdot 10^{12}\rm{cm}^{-2}$). The dominant superconducting instability in this regime has p--wave symmetry. Upon decreasing the density of holes, $\lambda$ increases gradually. Approaching the zero hole density limit, $\lambda$ increases sharply, and the dominant pairing channel changes to d--wave. $\lambda$ diverges in the limit $n_e \rightarrow 0$, due to the diverging density of states~\cite{SI}.

\emph{Kohn-Luttinger mechanism in RTG.---}Turning to RTG, we use a realistic model for the band structure and  $\Lambda_{\bm{k},\bm{q},\tau}$~\cite{SI}. The resulting FSs 
of the inner and outer pockets in Fig.~\refsub{fig1}{(e)} show a substantial trigonal warping, leading to anisotropic $\Pi_{0,\bm{q}}$. However, some of the features are similar to those of the circularly symmetric model. $\lambda$ as a function of density [Fig.~\refsub{fig1}{(g)}] is qualitatively similar to that of the circularly symmetric model, showing a sharp increase at the density where the inner pocket appears ($n_e\approx -2.1\cdot 10^{12} \rm{cm}^{-2}$), followed by a gradual increase of $\lambda$ with decreasing hole density. The symmetry of the order parameter changes from extended s--wave ($A_1$ representation of the point group $C_{3v}$ of RTG with a perpendicular electric field), with an opposite sign on the inner and outer Fermi pockets, to p--wave (representation $E$). 
$\lambda$ diverges logarithmically at the Van Hove density $n_e \approx -0.5\cdot 10^{12} \rm{cm}^{-2}$, where the annulus reconstructs into three disconnected pockets. Focusing on the p--wave phase and going beyond the linearized gap equation, the favored state below $T_c$ is a chiral $p_x + ip_y$ state, which can be derived by considering the quartic terms in the Ginzburg-Landau free energy functional~\cite{SI}. 

Importantly, in Eq~\eqref{eq:H}, spin singlet and triplet pairing are degenerate, independently of the orbital symmetry of the order parameter. This counter-intuitive result follows from the $SU(2)\times SU(2)$ symmetry of the Hamiltonian, which allows to rotate the spin of one valley relative to the other \cite{lee2019theory,Khalaf2020,Chou2021}. In the singlet (triplet) case, the order parameter in valley $\tau=+1$ has the same (opposite) phase to that of valley $-1$, such that the Pauli principle is obeyed. We shall discuss the lifting of the singlet-triplet degeneracy below.

\begin{figure}[t]
    \centering
    \includegraphics[width=0.99\columnwidth]{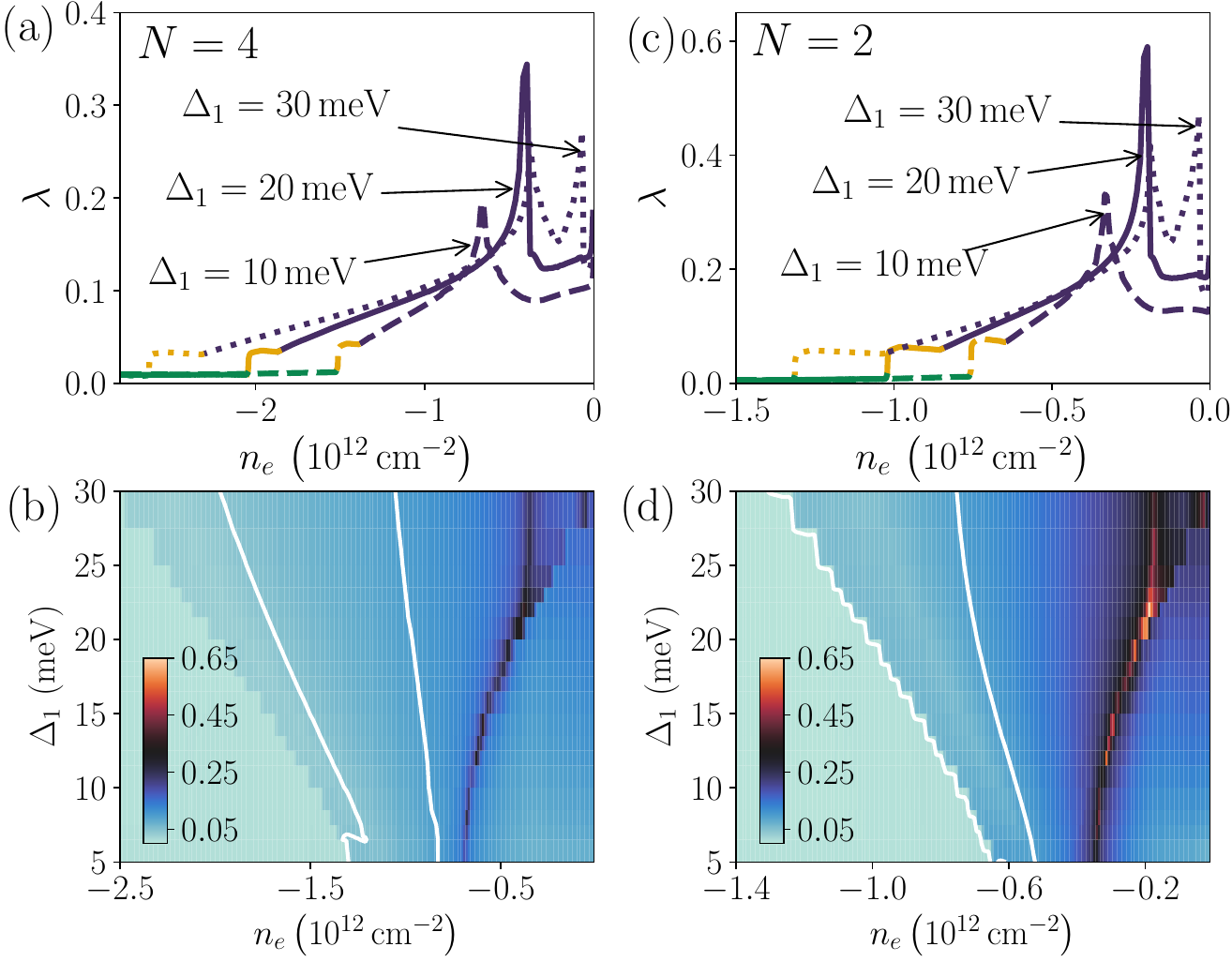}
    \caption{(a,b) Coupling constant $\lambda$ as a function of density and displacement field $\Delta_1$ for the unpolarized state, $N=4$. (c,d) Same as a,b assuming the spin polarization, $N=2$. The color coding in (a,c) is identical to Fig.~\ref{fig1}. The white contours in (b,d) indicate  $\lambda=0.05,\, 0.1$.
    }
    \label{fig2}
\end{figure}

Fig.~\refsub{fig2}{(a,b)} shows the density and $\Delta_1$ dependence of $\lambda$ 
in the spin and valley unpolarized phase [$N=4$ in Eq.~\eqref{eq:RPA}], corresponding to the SC1 region in experiment~\cite{Zhou2021superconductivity}. Increasing $\Delta_1$ shifts the peak corresponding to the Van Hove singularity (VHS) towards charge neutrality. However, there is a broad density regime ($n_e <-0.8\cdot 10^{12}\rm{cm}^{-2}$) where $\lambda$ is an increasing function of $\Delta_1$ at a fixed density, following the trend of the density of states with $\Delta_1$ in this regime. This is consistent with the fact that the SC1 phase is entered upon increasing electric field in the unpolarized phase~\cite{Zhou2021superconductivity}. To account for $T_c=0.1$ K in SC1, assuming that $W\approx E_F\approx 50$ K, we need $\lambda\approx 0.16$. As seen in Fig.~\ref{fig2}, our theory can produce $\lambda$s of this order.
In the vicinity of the VHS, $\lambda$ diverges logarithmically, and our theory breaks down. The VHS are not attained in experiment, since they are preceded by Stoner transitions~\cite{Zhou2021half}.
We note that the dispersion features a higher order VHS at $\Delta_1\approx18\,\mathrm{meV}$~\cite{SI,Yuan2019}.

Fig.~\refsub{fig2}{(c,d)} shows $\lambda$ as a function of density and $\Delta_1$ assuming a spin polarized, valley-unpolarized state ($N=2$), which corresponds to the experimentally observed normal state of SC2 \cite{Zhou2021half,Zhou2021superconductivity,comment_valley}. The overall trends are similar to those of Fig.~\refsub{fig2}{(a,b)}, although for a spin-polarized normal state, the pairing is necessarily spin triplet. The magnitude of $\lambda$ is somewhat larger than for $N=4$, due to stronger effective interaction [Eq.~\eqref{eq:RPA}]. This is in apparent disagreement with experiment. We speculate that the reason for this discrepancy is the residual Hund's coupling, not considered in Eq.~\eqref{eq:H}, which may favor spin singlet pairing (see below).

\begin{figure}[b]
    \centering
    \includegraphics[width=\columnwidth]{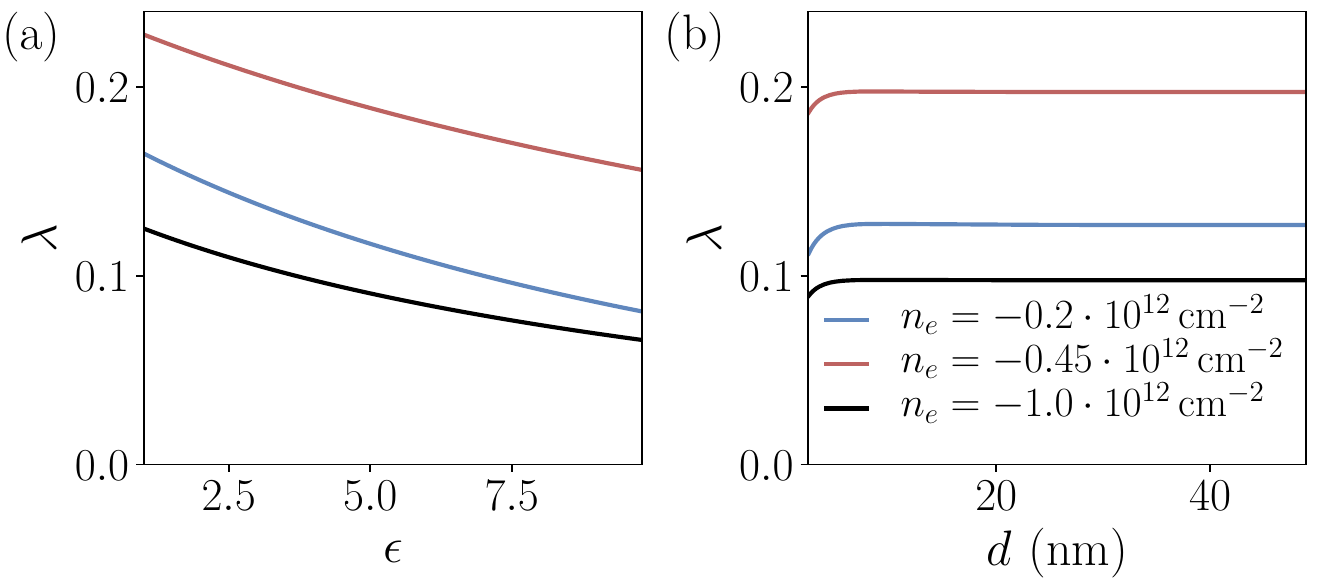}
    \caption{$\lambda$ as a function of dielectric constant $\epsilon$ (a) and distance to the gate $d$ (b). The results are for $\Delta_1=20\,\mathrm{meV}$ and densities listed in the legend correspond 
    to the three pocket FS (blue), two annular FS near VHS (red) and the regime of annulus away from the singularity (black).}  
    \label{fig3}
\end{figure}

Finally, in Fig.~\ref{fig3} we show the dependence of $\lambda$ on the dielectric constant $\epsilon$ and the distance to the gates, $d$, for three values of the electron density. $\lambda$ increases with decreasing $\epsilon$, since superconductivity originates from repulsive interactions. Interestingly, $\lambda$ is nearly independent of $d$ for $d\gtrsim 4\rm{nm}$ because the pairing benefits primarily from large momentum scattering processes where $q$ is of the order of $2k_F$~\cite{Note2}.

\emph{Role of the Hund's coupling.---}The lifting of the degeneracy between between singlet and triplet pairing requires an inter-valley interaction that depends on the relative spin of the two electrons. Such an interaction is short-ranged in real space, and receives contributions from the short-range part of the Coulomb interaction and from electron-phonon coupling~\cite{Alicea2006,Kharitonov2012}. The simplest term of this form is an inter-valley Hund's coupling, $H_{\rm{Hund}} = -J_H\int\,d^2r \bm{S}_+\cdot\bm{S}_-$, where $\bm{S}_{\pm}$ are the spin densities in the two valleys. $J_H$ may be of either sign.  

In general, $J_H>0$ ($J_H<0$) favors triplet (singlet) pairing, respectively. However, for e.g.~a $p_x+ip_y$ state, $H_{\rm{Hund}}$ drops out of the linearized gap equation (\ref{eq:lin}) and does not lift the singlet-triplet degeneracy~\cite{SI}. Physically, this is because $H_{\rm{Hund}}$ is local in real space, and  non-zero angular momentum of $p_x+ip_y$ leads to vanishing amplitude of the two electrons to be at the same location. 

To lift the degeneracy in the $p_x + ip_y$ state, one needs to go beyond the assumption of a momentum independent (i.e., local) Hund's term. We assume that the effective Hund's coupling depends on distance, $J_H(\bm{r}-\bm{r}')$. Expanding its Fourier transform $\tilde{J}_H$ at small momenta as: $\tilde{J}_H(\bm{q}) = J_{H,0} + J_{H,2}\,(qa)^2 + O(q^4)$ (where $a$ is the graphene lattice spacing), we find that ${J}_{H,2}>0$ (${J}_{H,2}<0$) favors singlet (triplet) pairing \cite{SI}. Hence, the lifting of the degeneracy is determined by the sign of the residual non-local Hund's coupling, which can be of either sign relative to $J_{H,0}$, depending on microscopic details. ${J}_{H,2}>0$ corresponds to a negative second moment of $J_H(\bm{r})$, i.e., the effective Hund's coupling is \emph{antiferromagnetic} in some range of distances, which promotes singlet pairing. 

Deriving the effective Hund's term microscopically is notoriously difficult~\cite{Kharitonov2012}. Instead, we can determine the signs of $J_{H,0}$ and $J_{H,2}$ from the experiment. We assume that $J_{H,0}>0$, favoring ferromagnetism, and $J_{H,2}>0$, favoring singlet pairing in the $p_x+ip_y$ phase. Hence, the SC1 phase, which emerges out of a spin-unpolarized normal state, is spin singlet. Note that for a simple s-wave order parameter (with no sign change of $\Delta_{\bm{k}}$ within a single valley), the leading-order $J_{H,0}$ term is sufficient to lift the singlet-triplet degeneracy. Hence, $J_{H,0}>0$ (required by the presence of valley-unpolarized ferromagnets) implies that SC1 -- if it were simple s-wave -- should be spin triplet, in contradiction with experiment. 

The normal state of SC2 is a fully spin polarized ferromagnet, and hence SC2 must be spin triplet, which is consistent with the strong violation of the Pauli limit in SC2 seen in experiment. In our model, the spin triplet state is disfavored by $J_{H,2}>0$, which can potentially explain the fact that $T_c$ of SC2 is substantially lower than that of SC1 \cite{Zhou2021superconductivity}. 
That being said, we emphasize that for SC2,  spin triplet pairing is still much favored over spin singlet, because there are no opposite spins to pair with at the Fermi energy. 

\emph{Discussion.---}We demonstrated that repulsive interactions can give rise to robust unconventional superconductivity in two-dimensional systems with annular FSs that are similar to RTG in the parameter regimes where superconductivity was recently discovered. The Kohn--Luttinger mechanism is consistent with the absence of a strong temperature dependence of the resistivity above $T_c$ 
and can further explain the apparent discrepancy between SC1 being spin singlet and the presence of a nearby spin polarized phase, implying that the inter-valley exchange coupling is ferromagnetic.

An outstanding question is whether the superconducting state in RTG is indeed unconventional. In contrast to s-wave pairing~\cite{Anderson1959theory,Finkelstein1994suppression}, unconventional SC is very sensitive to non-magnetic disorder~\cite{Abrikosov,Larkin,Millis1988,Radtke1993}. Thus, introducing controlled amounts of disorder to the system could be used to identify the nature of the superconducting state~\cite{Mackenzie1998}. 
Another 
prediction of our theory is the near-degeneracy between singlet and triplet pairing, which is also expected  within an electron-phonon mechanism~\cite{Chou2021}. As SC1 is suppressed by an in-plane field, a triplet state may be stabilized at lower temperature. So far, $T_c$ of SC1 was found to be suppressed below $50$mK with the application of an in-plane field \cite{Zhou2021superconductivity}, but a triplet SC with a lower $T_c$ is possible. Furthermore, a fully spin polarized SC2 is expected to have exotic properties, due to the intertwining of the SC phase with the orientation of the magnetization~\cite{Cornfeld2021}. The most likely pairing state within our theory is a chiral $p_x + ip_y$ SC that breaks time reversal symmetry and generates spontaneous edge currents~\cite{Furusaki2001,Stone2004,Sauls2011}. 
Since within the symmetry group of RTG p--wave and d--wave pairings are not distinct from each other, the obtained pairing is reminiscent of the spin-singlet chiral d--wave pairing predicted for doped graphene \cite{Black2014chiral,Black2007resonating,Wu2013correlated,Nandkishore2012chiral,Wang2012functional,Kiesel2012competing}.  

The pairing mechanism proposed in this work is mediated by electronic fluctuations with a broad spectrum, and does not assume the existence of a soft collective mode at low energies (that arises if the system is close to a continuous phase transition). If such soft mode exists, it could enhance $T_c$ further. 
Very recent studies of this scenario~\cite{Chatterjee,Levitov} found a chiral p-wave state similar to our work.

Finally, it is desirable to raise $T_c$ in RTG and other systems with annular FSs. Our theory predicts a significant enhancement of $\lambda$ near the VHS where the inner and outer FSs meet. In RTG, the metallic state becomes unstable towards spin or valley polarization before this point is reached. We speculate that decreasing distance to metallic gates $d$ may suppress these instabilities, while not affecting superconductivity significantly until $d\approx 4$nm (Fig. \ref{fig3}), effectively enhancing SC phase. 

\begin{acknowledgments}
We thank Yang-Zhi Chou, Andrey Chubukov, Johannes Hofmann,
Steve Kivelson, Sri Raghu, 
and Sankar das Sarma, Jay Sau, Fengcheng Wu, and Andrea Young for many stimulating 
discussions and for their comments on the manuscript. E.B. thanks S. Chatterjee, T. Wang, and M. Zaletel for a collaboration on a related topic. A.G. acknowledges support by the European Unions Horizon 2020 research and innovation program under the Marie Sklodowska-Curie Grant Agreement No.~754411. E.B. and T.H. were supported by the European Research Council (ERC) under grant HQMAT (Grant Agreement No.~817799), by the Israel-USA Binational Science Foundation (BSF), and by a Research grant from Irving and Cherna Moskowitz.

\end{acknowledgments}

%

\renewcommand{\thefigure}{S\arabic{figure}}
\renewcommand{\figurename}{Figure}
\setcounter{figure}{0}
\setcounter{equation}{0}
\renewcommand{\theequation}{S\arabic{equation}}

\clearpage

\onecolumngrid

\begin{center}
	\textbf{\large Supplementary material for: \strut\\ ``\mytitle'' }\\[5pt]
\end{center}

\section{Band structure and wave functions of rhombohedral trilayer graphene}

The band structure which is used to calculate superconducting pairing is based on $6\times6$ model \cite{Koshino2009,MacDonald2010,Zhou2021half}
\begin{equation}
h_{\bm{k}}=\left(\begin{array}{cccccc}
\delta+\Delta_{1}+\Delta_{2} & v_{0}\pi^{*} & v_{4}\pi^{*} & v_{3}\pi & 0 & \frac{\gamma_{2}}{2}\\
v_{0}\pi & \Delta_{1}+\Delta_{2} & \gamma_{1} & v_{4}\pi^{*} & 0 & 0\\
v_{4}\pi & \gamma_{1} & -2\Delta_{2} & v_{0}\pi^{*} & v_{4}\pi^{*} & v_{3}\pi\\
v_{3}\pi^{*} & v_{4}\pi & v_{0}\pi & -2\Delta_{2} & \gamma_{1} & v_{4}\pi^{*}\\
0 & 0 & v_{4}\pi & \gamma_{1} & -\Delta_{1}+\Delta_{2} & v_{0}\pi^{*}\\
\frac{\gamma_{2}}{2} & 0 & v_{3}\pi^{*} & v_{4}\pi & v_{0}\pi & \delta-\Delta_{1}+\Delta_{2}
\end{array}\right),
\label{eq:6by6Ham}
\end{equation}
where $\pi=\tau k_x+ik_y$ ($\tau$ is the valley index) and the Hamiltonian is written in the basis $(A_1,B_1,A_2,B_2,A_3,B_3)$ with $A_i$ and $B_i$ denoting different sublattice sites on layer $i$. Here the $v_i=\sqrt{3}a\gamma_i/2$, $a=2.46\AA$ is the lattice constant of graphene, $\Delta_1$ is proportional to the perpendicular electric field, and $\gamma_i$, $\delta$, $\Delta_2$ are the fixed material parameters: $\gamma_0=3.1\,\mathrm{eV}$, $\gamma_1=0.38\,\mathrm{eV}$, $\gamma_2=-0.015\,\mathrm{eV}$, $\gamma_3=-0.29\,\mathrm{eV}$, $\gamma_4=-0.141\,\mathrm{eV}$, $\delta=-0.0105\,\mathrm{eV}$, $\Delta_2=-0.0023\,\mathrm{eV}$\cite{Zhou2021half}. 

\begin{figure}[b]
    \centering
    \includegraphics[width=0.8\columnwidth]{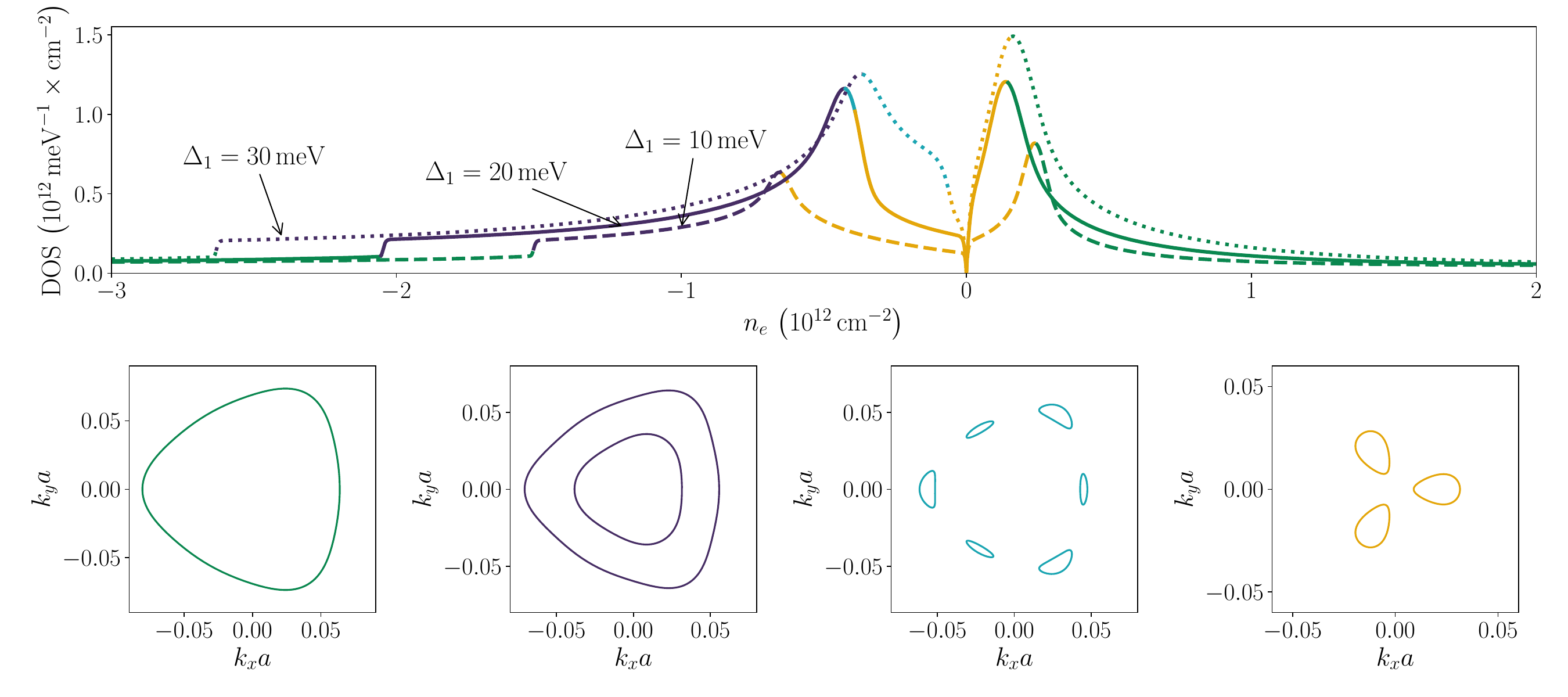}
    \caption{Density of states of $6\times6$ band model (\ref{eq:6by6Ham}) for three different values of the displacement field $\Delta_1$. Different regions of density corresponding to different FS topology are colored differently and representative FSs are shown in the lower panel. The FSs are for $\Delta_1=30\,\mathrm{meV}$ and $n_e=-2.63\times10^{12}\,\mathrm{cm^{-2}}$, $n_e=-1.35\times10^{12}\,\mathrm{cm^{-2}}$, $n_e=-0.13\times10^{12}\,\mathrm{cm^{-2}}$, $n_e=0.14\times10^{12}\,\mathrm{cm^{-2}}$, respectively. The DOS data is generated at the temperature $T=0.025\,\mathrm{meV}$ which regularizes VHSs.}  
    \label{fig:dos}
\end{figure}

\begin{figure}[t]
    \centering
    \includegraphics[width=0.6\columnwidth]{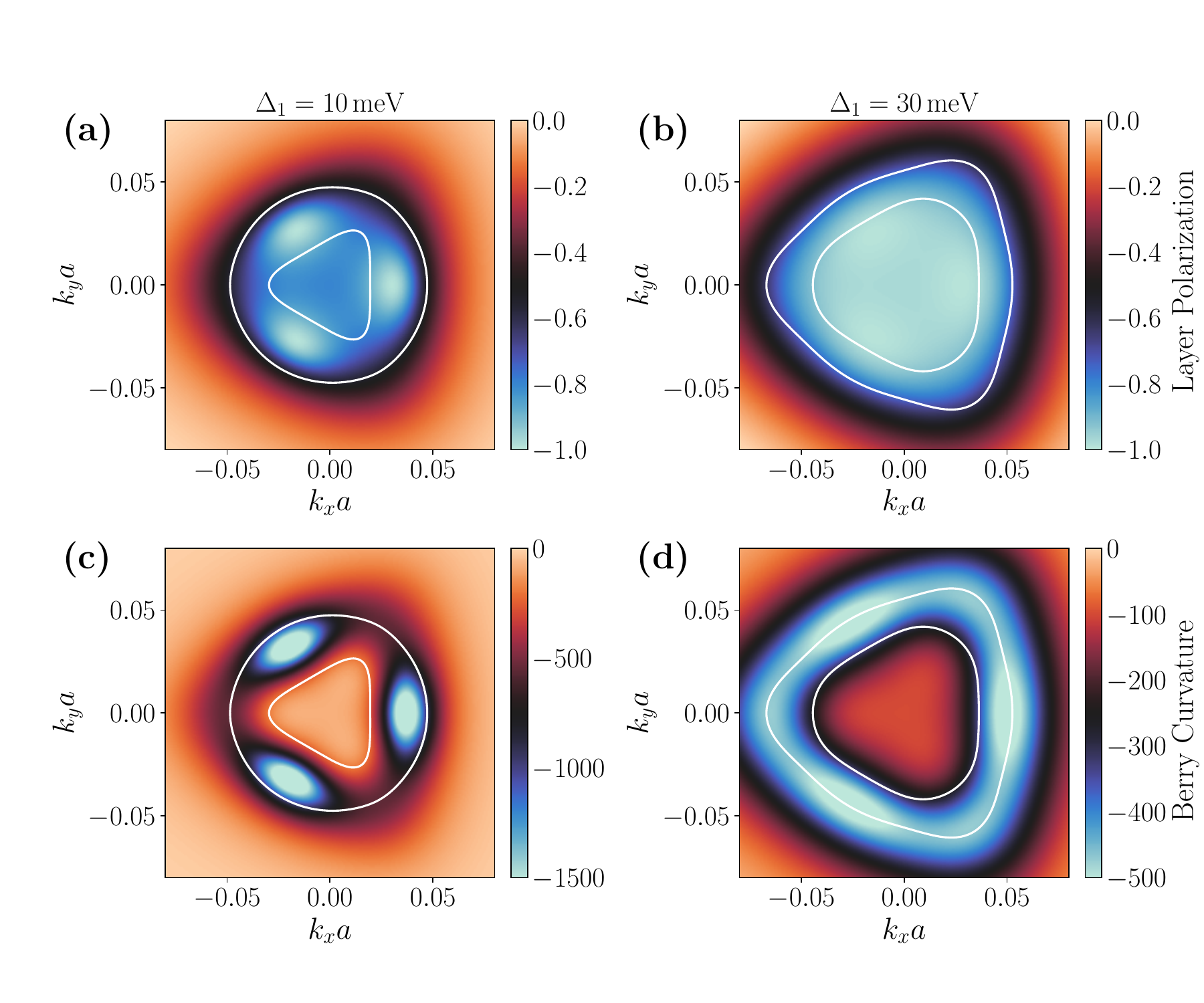}
    \caption{Evolution of layer polarization (a-b) and Berry curvature (c-d) with displacement field. The wave functions become more layer polarized upon increasing $\Delta_1$.
    First column shows layer polarization (a) and Berry curvature (c) for  $\Delta_1=10\,\text{meV}$; second column corresponds to  $\Delta_1=30\,\text{meV}$. While lines show FS corresponding to fixed density $n=-0.9\cdot 10^{12}\,\text{cm}^{-2}$ in both cases.}
    \label{fig:spin}
\end{figure}

Fig.~\ref{fig:dos} shows the density of states (DOS) of the $6\times6$ model. Generally the system supports three types of Van Hove singularities (VHS): (1) disappearance of the central pocket for large (negative) hole densities, (2) transition from three pockets into six pockets (this regime is realized only for $\Delta_1\gtrsim 18\,\mathrm{meV}$); and (3) merging of three or six pocket into annular or single contour.  
We note that at the displacement field $\Delta_1\approx 18\,\mathrm{meV}$ where the phase with six pockets terminates, the dispersion features a higher order VHS. It is of the same type which was discussed recently for twisted bilayer graphene~\cite{Yuan2019}. At the higher order VHS, the density of states diverges more strongly compared to a normal VHS. Annular FSs are present mostly for the hole densities and are responsible for the unconventional superconductivity discussed in the main text. In contrast, on the electron side, the annular Fermi surface is present for a very narrow range of chemical potentials ($\Delta\mu\leq 0.01\,{\rm meV}$ for $\Delta_1\leq 20\,{\rm meV}$), it emerges once the three pockets merge at the VHS and rapidly disappears.

\begin{figure}[b]
    \centering
    \includegraphics[width=\columnwidth]{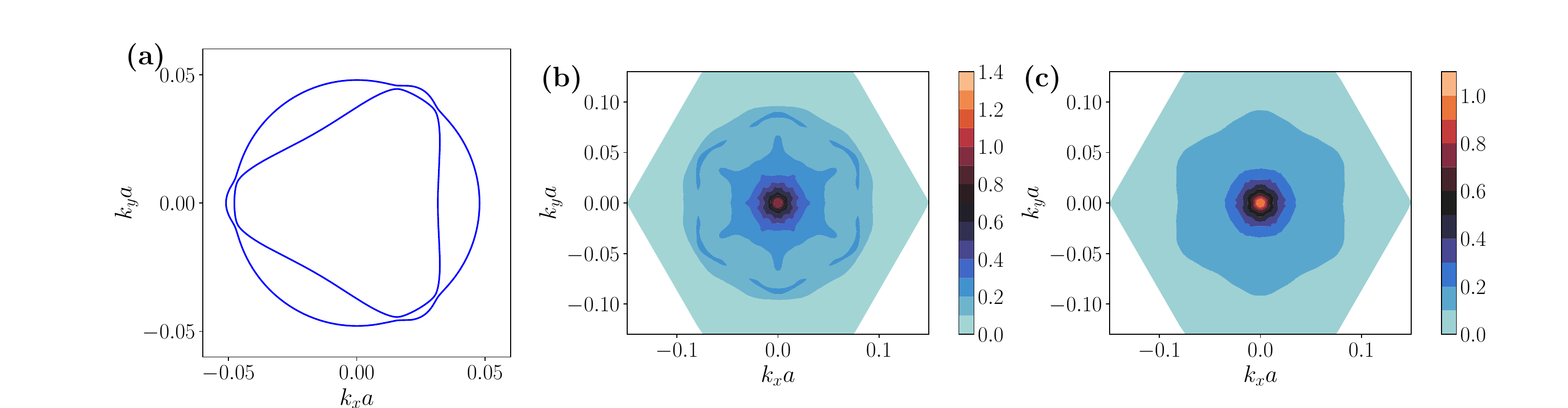}
    \caption{(a) Fermi surface of $6\times6$ band structure close to VHS. (b) Static polarization function $\Pi_{0,\bm{q}}$ without Bloch wave functions, and full polarization function (c) have similar structure. The $\Pi_{0,\bm{q}}$ is normalized by the density of states so that $\Pi_{0,0}=1$. The additional peaks in (b) are not visible in (c), due to the structure of wave functions. The results are for $\Delta_1=20\,\mathrm{meV}$ and $n_e=-0.42\times10^{12}\,\mathrm{{cm}^{-2}}$. Finite temperature $T=0.015\,\mathrm{meV}$ is used in polarization calculation to smoothen the singularities.}
    \label{fig:chi-2d}
\end{figure}

In addition to its complex Fermi surfaces, the $6\times6$ model differs from the simplified circularly symmetric model in that it incorporates the Bloch wave functions into the static polarization function and directly into the linearized gap equation. To understand the role of the wave functions in the calculations we show layer polarization and Berry curvature as a function of $\bm{k}$ in Fig.~\ref{fig:spin}. At low energies, the $6\times6$ model can be approximated by a $2\times2$ model, where only $A_1$ and $B_3$ sublattice sites contribute \cite{MacDonald2010}. The layer polarization is defined as $(n_1 - n_3)/(n_1 + n_3)$, where $n_{1,3}$ are the densities in the first and third layers, respectively. 
While for densities corresponding to annular Fermi surfaces the wave function is strongly layer polarized, it still features three peaks both in polarization and Berry curvature figures.  
Increasing the displacement field $\Delta_1$ enhances the layer polarization, thereby decreasing the influence of the wave function. 

Finally, Fig.~\ref{fig:chi-2d} shows the static polarization function $\Pi_{0,\bm{q}}$ close to the VHS point, both with and without taking into account the form factors $\Lambda_{\bm{k},\bm{q},\tau}$ due to the projection on a single band. The inclusion of the wave functions suppresses the additional peaks of the static polarization at finite $\bm{q}$. This results in an enhancement of effective interaction within RPA. While this by itself would increase the values of $\lambda$ obtained in the calculation, one has to keep in mind that the wave functions also enter directly into the linearized gap equation [cf. Eq.~(4) in the main text]. These two effects tend to cancel each other, making the combined effect from the inclusion of the wave functions not substantial for the amplitude of $\lambda$.

\section{Method of calculating $T_c$}

\begin{figure}[t]
    \centering
    \includegraphics[width=.49\columnwidth]{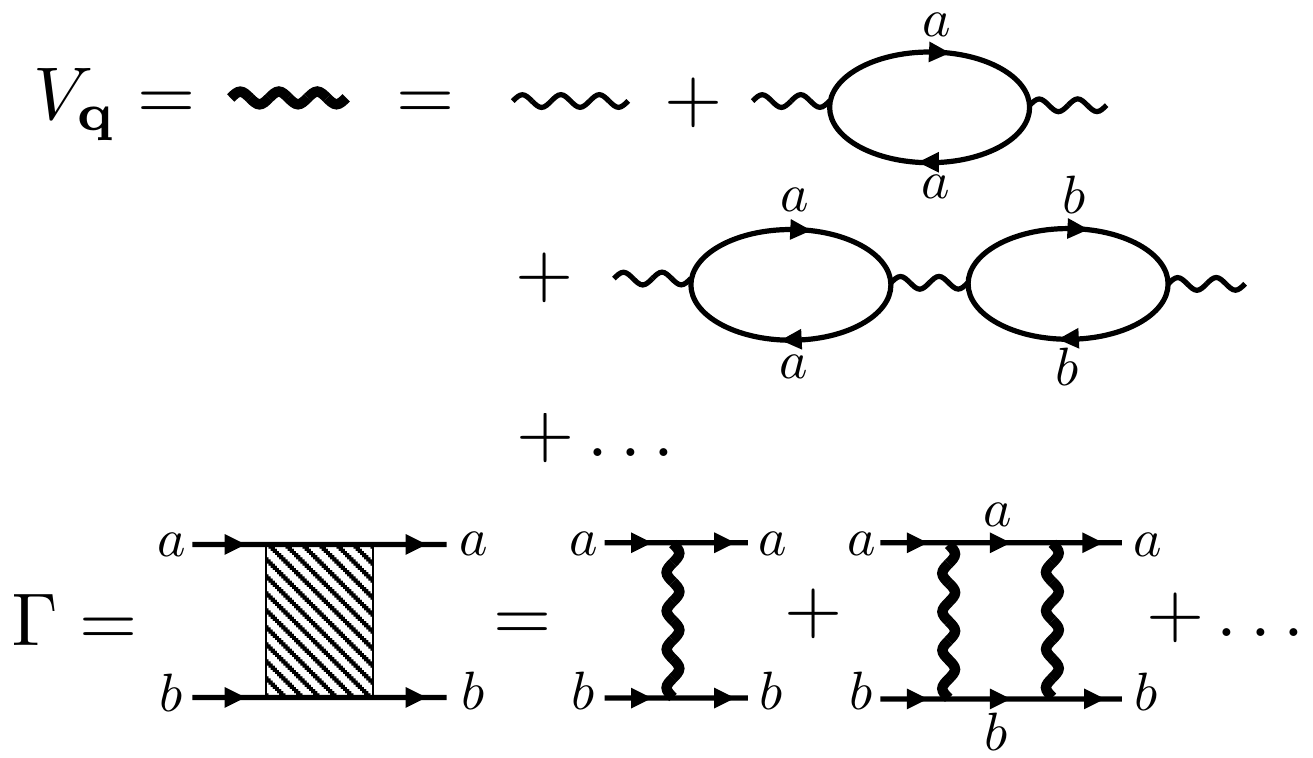}
    \caption{Diagrammatic representation of the effective interaction $V_{\bm{q}}$ and the pairing vertex $\Gamma$. Thin wiggly lines denote the Coulomb interaction, solid lines are fermionic propagators with flavor index $a,b$.}
    \label{fig:RPA}
\end{figure}

Here, we provide more details of our method of treating the superconductivity emergent from the Hamiltonian (\ref{eq:H}). To formally control the calculation, we imagine taking the limit of large number of electron flavors, $N\rightarrow \infty$ (with $N/2$ flavors in each of the two valleys). At the same time, we take the weak coupling limit such that $N e^2/\epsilon \rightarrow \rm{const}$. To leading order in $1/N$, the effective interaction is then given by the RPA expression, Eq.~(3) in the main text (see Fig.~\ref{fig:RPA}). 

We can then use this interaction to calculate the pairing vertex $\Gamma$ for inter-valley pairs (Fig.~\ref{fig:RPA}). Within our approach, the effective pairing interaction for a given flavor is of order $1/N$, which justifies the weak coupling treatment. Each additional term in the ladder summation is suppressed by $1/N$ but enhanced by a BCS logarithmic factor, of order $\log(W/T)$, and hence we need to sum the infinite ladder series, giving the usual BCS gap equation. Applying weak-coupling BCS theory is reasonable when making connection to experiments in RTG, since $T_c/E_F \sim 10^{-3}$ is much smaller than 1. 

Note that our setup of the problem does not give the instabilities towards flavor symmetry breaking half and quarter metals seen in RTG; in the limit in which our calculations are justified, these effects are suppressed by factors of $1/N$. Treating superconductivity and flavor symmetry breaking on the same footing in a theoretically controlled way is difficult, and goes beyond our present analysis. On physical grounds, however, it is reasonable that upon changing the hole density towards the Van Hove point, superconductivity is first enhanced due to the increase in the density of states. By the Stoner criterion, the same increase in the density of states promotes flavor symmetry breaking transition that decreases the density of states, suppressing superconductivity. Hence, superconductivity and symmetry breaking appear close to each other in the phase diagram, even though they are not necessarily causally related. 

\subsection*{Linearlized gap equation}

To derive Eq. (4) of the main text, we start from the standard BCS mean-field equation for the gap function:
\begin{equation}
    \Delta_{\bm{k}}=-\int\frac{d^{2}k'}{(2\pi)^{2}}V(\bm{k}-\bm{k}')\left|\Lambda_{\bm{k},\bm{k}'-\bm{k},+1}\right|^{2}\frac{\tanh\left(\frac{E_{\bm{k}',1}}{2T}\right)}{2E_{\bm{k}',1}}\Delta_{\bm{k}'},
    \label{eq:lin_sup}
\end{equation}
where $E_{\bm{k},\tau}=\sqrt{\varepsilon_{\bm{k},\tau}^{2}+|\Delta_{\bm{k}}|^{2}}$ is the dispersion of Bogoliubov quasi-particles. To get $T_c$, we keep only terms to linear order in $\Delta_{\bm{k}'}$ in the right hand side of \eqref{eq:lin_sup}, replacing $E_{\bm{k},1}$ by $|\varepsilon_{\bm{k},1}|$. Furthermore, we linearize the dispersion near the FS, and change variables to Fermi surface coordinates such that $\int \frac{d^2 k'}{(2\pi)^2} = \int \frac{dk'_\parallel\,dk_\perp}{(2\pi)^2}$, where $k_\parallel$ and $k_\perp$ are momenta parallel and perpendicular to the FS, and $\varepsilon_{\bm{k},1} \approx v_{\bm{k}} k_\perp$. Integrating over $k_\perp$ gives a factor of $\frac{1}{v_{\bm{k}'}} \log(W/T)$, where $W$ is an energy cutoff of the order of $E_F$. This gives Eq. (4).

\subsection*{Solving linearized gap equation numerically}

\begin{figure}
    \centering
    \includegraphics[width=.99\columnwidth]{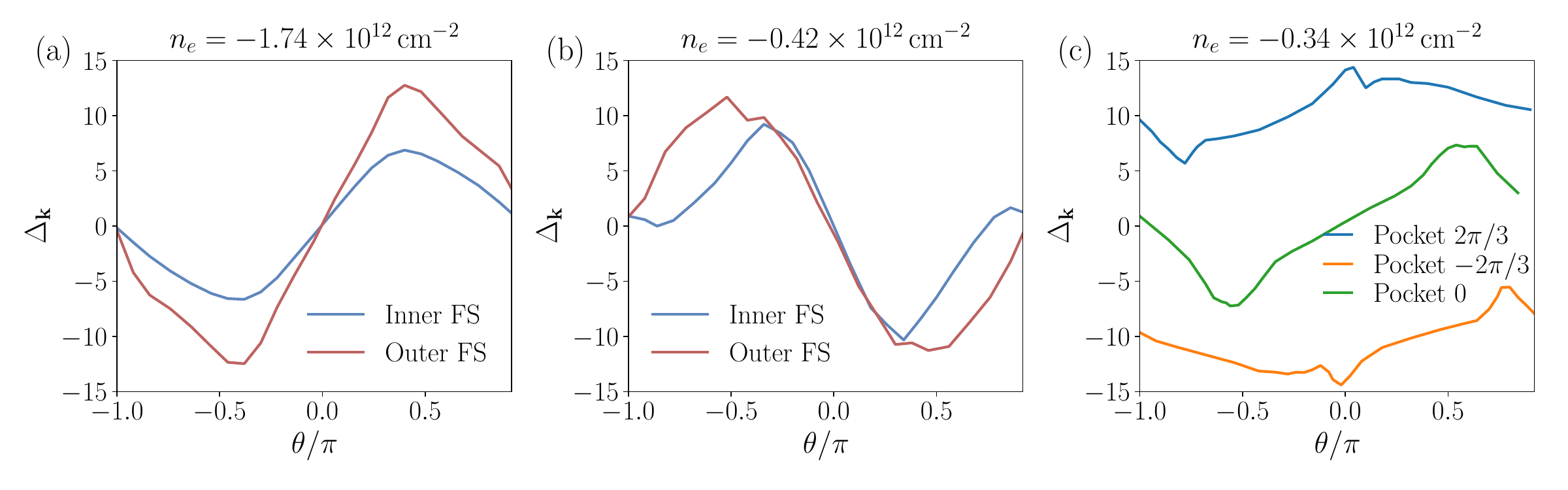}
    \caption{
    Angular dependence of the superconducting order parameter along the Fermi surface, for a p-wave state with a node along the x-axis (cf. Fig.~1 in the main text). (a) At large density, the inner pocket is small and the order parameter has a smaller amplitude there. (b) When the annulus is thin and both Fermi surfaces approach each other, the order parameters have comparable magnitude. (c) In the regime with three pockets, only the order parameter in the pocket along the nodal direction has a sign change, with the other two being of opposite sign with respect to each other. We note that the overall magnitude of the order parameter cannot be determined from the linearized gap equation and is fixed using normalization condition.
    }
    \label{fig-wf}
\end{figure}

In this section we give the details of the numerical evaluation of $\lambda$ and of the superconducting order parameter. To solve the linearized gap equation (\ref{eq:lin}) numerically we first calculate the static polarization $\Pi_{0,\bm{q}}$ by taking a discrete square grid of momenta $ka$ around the $K$ and $K^\prime$ point in the range of $[-0.18,0.18]$ with $2000$ points in each direction. The static polarization is calculated at a finite temperature $T=0.015\,\mathrm{meV}$ to regularize the singularities of the polarization function. For each value of the density we choose angular equidistant points on the Fermi surface and calculate for those points the symmetric matrix
\begin{equation}
\widetilde{\mathcal{M}}_{\bm{k},\bm{k^\prime}}=-\frac{1}{\left(2\pi\right)^2}\sqrt{\frac{\Delta k \Delta k^\prime}{v_{\bm{k}}v_{\bm{k}^\prime}}}
V_{\bm{k}-\bm{k}^\prime}\left|\langle u_{\bm{k},+1}|u_{\bm{k}^\prime,+1}\rangle\right|^2,
\end{equation}
which is the discretized version of the linear operator defined in the main text. $\Delta k$ is the distance between two points on the Fermi surface. The number of points on the Fermi surface is chosen such that $\Delta k$ is approximately $0.25$ of the mean radius of the pocket but not smaller than 5 discrete grid points to regularize the divergence of $\lambda$ when density of states diverges. The largest eigenvalue of $\widetilde{\mathcal{M}}_{\bm{k},\bm{k}^\prime}$ corresponds to $\lambda$ presented in the main text. The superconducting order parameter is given by $\Delta_{\bm{k}}=\phi_{\bm{k}}\sqrt{v_{\bm{k}}/\Delta k}$, where $\phi_{\bm{k}}$ is the eigenvector of $\widetilde{\mathcal{M}}_{\bm{k},\bm{k^\prime}}$, i.e. it fulfills $\widetilde{\mathcal{M}}_{\bm{k},\bm{k^\prime}}\phi_{\bm{k}^\prime}=\lambda\phi_{\bm{k}}$.
Some examples for the superconducting order parameter calculated this way are shown in Fig.~\ref{fig-wf}. 

\subsection*{Superconductivity in the conduction band}
\begin{figure}[t]
    \centering
    \includegraphics[width=.79\columnwidth]{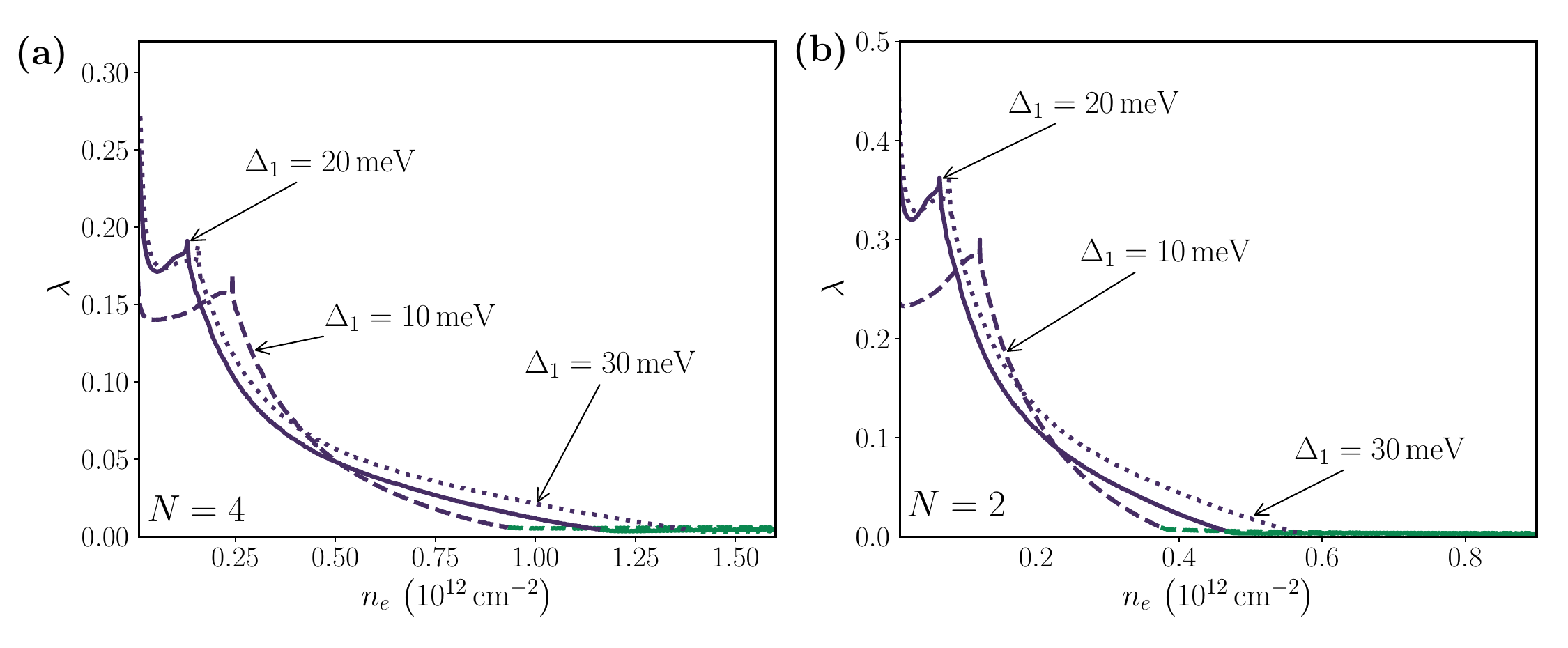}
    \caption{
    Coupling constant $\lambda$ as a function of electron density for different values of the displacement field $\Delta_1$ projected into the lowest electron band. (a) for the unpolarized $N=4$ state and (b) for the polarized $N=2$ state. In the broad range of densities the emergent SC order parameter is p-wave like (purple), while at larger densities it has 8 nodes and resembles g-wave (green). Since both p- and g-wave belong to the same representation of $C_{3v}$, the change in the nature of the order parameter happens via an avoided crossing of the largest eigenvalues.
    }
    \label{fig-le}
\end{figure}
We comment briefly on superconductivity on the electron side, where an annular Fermi surface is not realized for the system parameters of RTG. Fig.~\ref{fig-le} shows the dependence of the coupling constant $\lambda$ on the electron density for different values of the displacement field $\Delta_1$. While a p-wave superconducting instability is realized for this case as well, the presence of only a single Fermi pocket after the three small pockets have merged results in a more rapid decrease of $\lambda$ with increasing electron density. This naturally leads to smaller superconducting regions or absence of thereof, consistent with experimental data~\cite{Zhou2021superconductivity}. Near VHS a Stoner transition occurs, effectively making the region with a large density of states inaccessible. Thus, the more rapid decrease of $\lambda$ away from VHS observed for the conduction band gives rise to a smaller range of densities where $T_c$ is within reach of experimentally achievable temperatures. 

\section{Susceptibility and pairing in the circularly symmetric model}
In the following, we consider the more convenient parametrization of the ``Mexican hat dispersion'', $\epsilon_{\bm{k}}=\varepsilon_0((k^2- k_0^2)^2-\bar\mu)/k_0^4$. This differs from the one used in the main text by an overall minus sign, and uses different units for the chemical potential $\bar\mu=-k_0^4\mu/\varepsilon_0$. 
This choice of units highlights the two natural parameters that are present for this dispersion, the overall scale $\varepsilon_0$, and the size of the annulus as given by $k_0$.
The locations of the FSs are given by $k_{F1}^2=k_0^2+\sqrt{\bar\mu}$ and $k_{F2}^2=k_0^2-\sqrt{\bar\mu}$, the total number density is $n=N\sqrt{\bar\mu}/2\pi$. We will further use the abbreviation $q'=\sqrt{4k_0^2-q^2}$.
At $T=0$, the susceptibility can be written as
\begin{equation}
    \Pi_{0,\bm{q}}=
    N\int\frac{d^2k}{4\pi^2}
    \frac{\Theta(-\varepsilon_{\bm{k}})-\Theta(-\varepsilon_{\bm{k+q}})}{\varepsilon_{\bm{k+q}}-\varepsilon_{\bm{k}}}\\
    =\frac{N}{4\pi^2}\int_{k_{F2}}^{k_{F1}}\!\! kdk\int_{-\pi}^{\pi}\!\!d\theta
    \biggl(\frac{1}{\varepsilon_{\bm{k+q}}-\varepsilon_{\bm{k}}}-
    \frac{1}{\varepsilon_{\bm{k}}-\varepsilon_{\bm{k-q}}}
    \biggr).
\end{equation}
The angular integral is relatively straightforward, but as usual requires some care with regards to branch cuts in logarithmic functions. A bit more subtle is that a $\delta$-function appears in the radial dependence, a fact which is confirmed by a numerical integration of the regulated expression. 
After angular integration we obtain the following expression:
\begin{align}
    \Pi_{0,\bm{q}}&=\frac{N}{4\pi^2}\frac{k_0^4}{\varepsilon_0}
    \int_{k_{F2}}^{k_{F1}}\!\!     
    \frac{4 kdk}{q(k^2-k_0^2)}
    \Re\Biggl[
    \frac{i\pi}{2}\frac{q\sgn(4k_0^2-2q^2-4k^2)}{\sqrt{8k_0^2k^2-4k^4-q'^4}}
    \notag\\&\quad
    -\frac{\arctanh\frac{q+2k}{\sqrt{4k^2-q^2}}+\arctanh\frac{q-2k}{\sqrt{4k^2-q^2}}}{\sqrt{4k^2-q^2}}
    +\frac{\pi^2(k^2-k_0^2)}{4\sqrt{4k^2-q^2}}\delta(k-k_0)
    \Biggr].
\end{align}
This is simple enough to perform the radial integration. Surprisingly, the contribution from the $\delta$-function allows to combine the logarithms from the lower and upper boundary into one expression by canceling the additional contribution that is left from switching between branch cuts. This greatly simplifies the final result into
\begin{align}
    \Pi_{0,\bm{q}}&=-\frac{N}{4\pi^2}\frac{k_0^4}{\varepsilon_0}\Re\Biggl[
    \frac{2\pi }{qq'}
    \arctan\frac{\sqrt{4\bar\mu-2k_0^2q^2+q^4}}{qq'}\Theta(k_{F1}-k_{F2}-q)
    +\frac{i\pi}{q q'}\log\frac{2\sqrt{\bar\mu}+q'^2+q'\sqrt{q'^2+4\sqrt{\bar\mu}}}{2\sqrt{\bar\mu}-q'^2-q'\sqrt{q'^2-4\sqrt{\bar\mu}}}
    \Biggr].\label{eq:radsymmain}
\end{align}
For completeness, we also present the susceptibility for $\bar\mu>k_0^4$ (i.e. a single large Fermi circle after the annulus closes in the middle) which coincides with the solution for a dispersion where $k_0^2<0$ (i.e. a single large Fermi circle due to a positive mixed term in the dispersion). In these latter cases, it is
\begin{align}
    \Pi_{0,\bm{q}}&=-\frac{N}{4\pi^2}\frac{k_0^4}{\varepsilon_0}\Re\Biggl[
    \frac{\pi }{qq'}
    \arctan\frac{\sqrt{4\bar\mu-q^2q'^2}}{qq'}
    +\frac{\pi }{qq'}\arctan\frac{2k_0^2-q^2}{qq'}
    +\frac{i\pi}{q q'}\log\frac{k_0^2(q'^2+2\sqrt{\bar\mu}+q'\sqrt{q'^2+4\sqrt{\bar\mu}})}{\sqrt{\bar\mu}(q^2-2k_0^2-i qq')}
    \Biggr],\label{eq:radsymmain2}
\end{align}
where, as mentioned $k_0^2$ can assume values smaller than zero.

\begin{figure}
    \centering
    \includegraphics[width=.49\columnwidth]{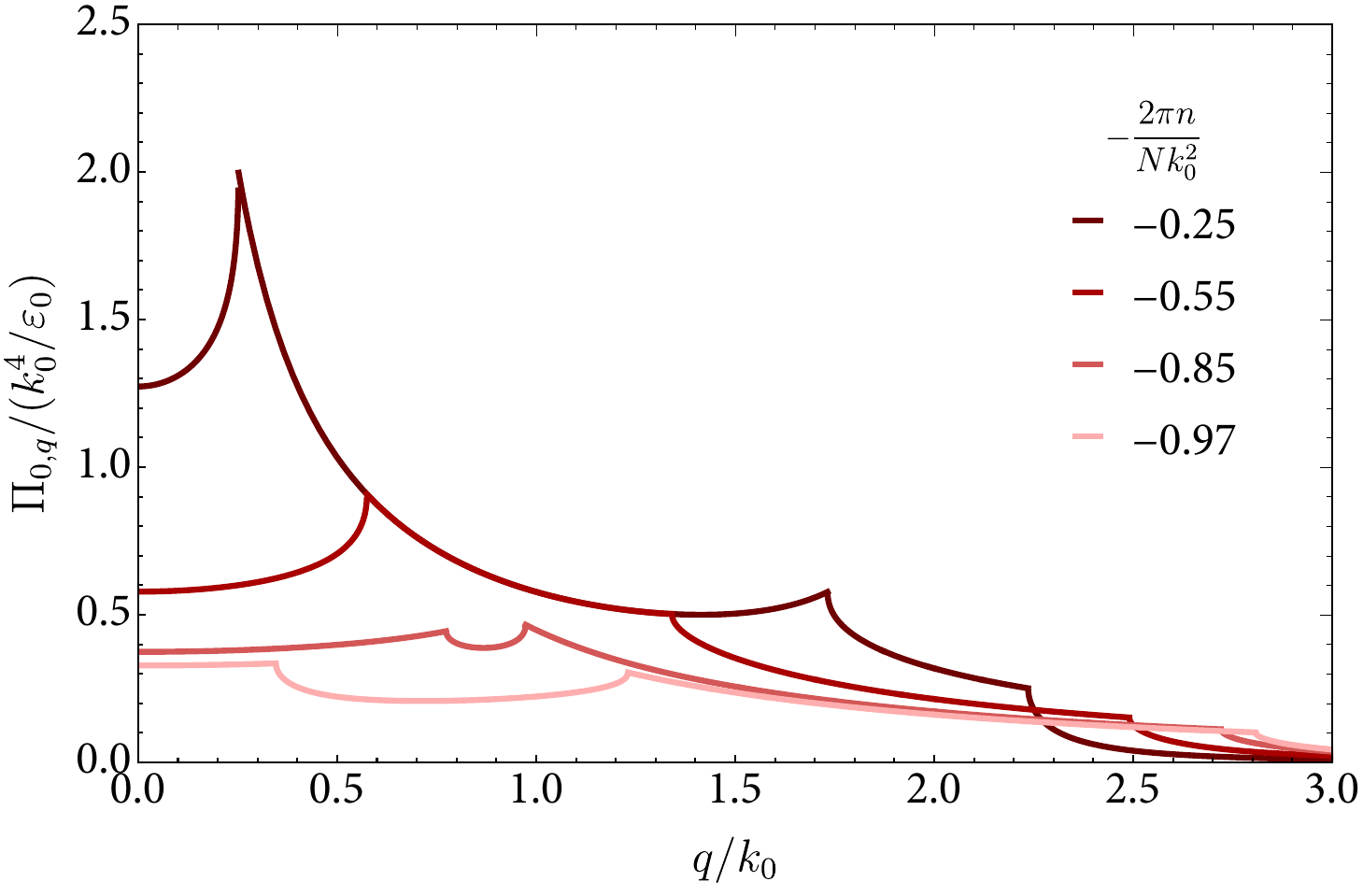}
    \includegraphics[width=.49\columnwidth]{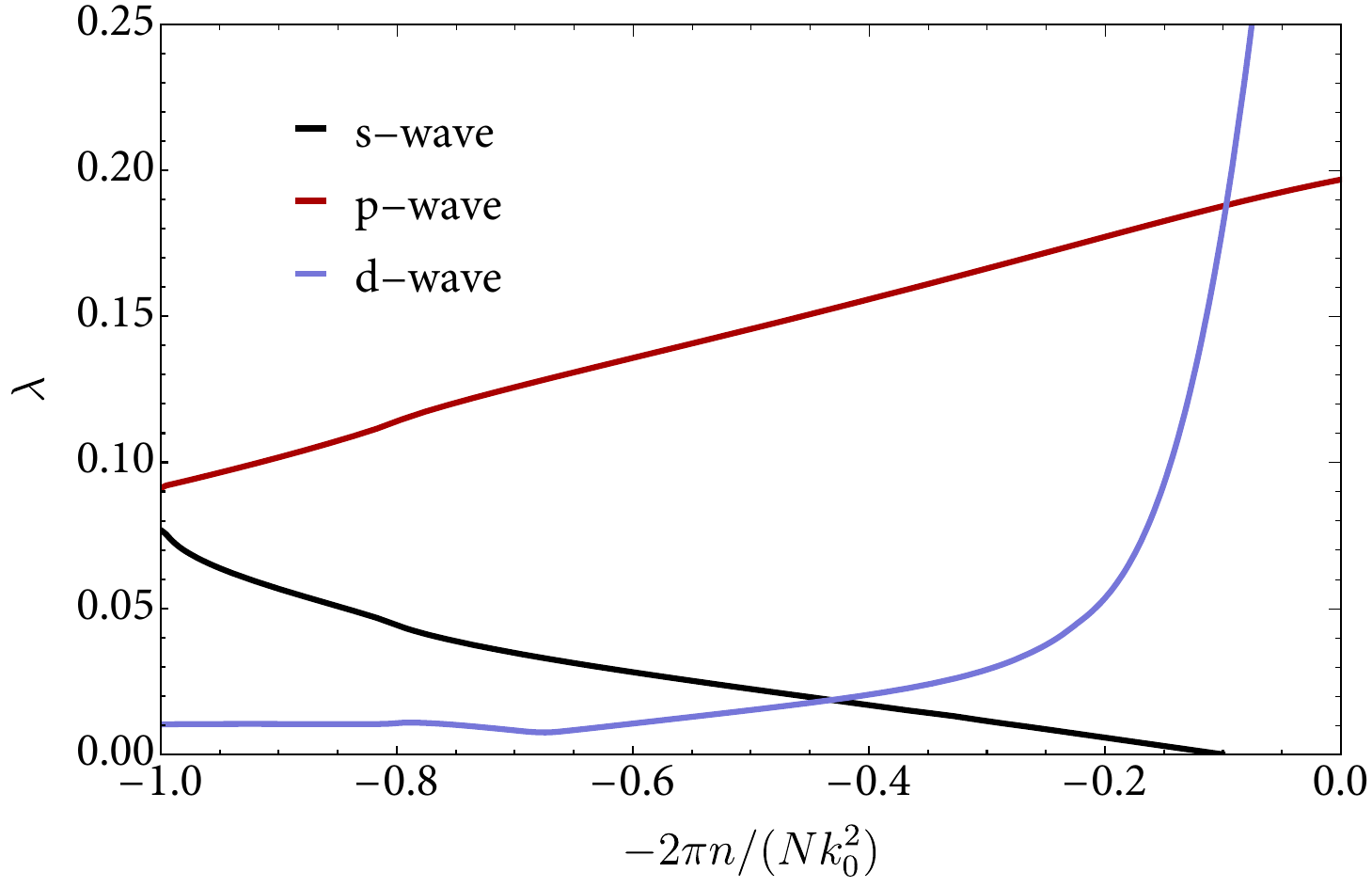}
    \caption{Mexican hat dispersion. (a) Dimensionless particle-hole susceptibility $\Pi_{0,\bm{q}}$ as a function of the normalized momentum transfer $q/k_0$. (b) Strong interaction limit $V_{\bm{q}}=\Pi_{0,\bm{q}}^{-1}$. Shown are the absolute value of the most negative eigenvalues in the s,p,d angular momentum channels as a function of normalized density. Both figures are for a number of flavors $N=4$, the total number density is $n$.}
    \label{fig:MH_lambda}
\end{figure}

We can use expression Eq.~\eqref{eq:radsymmain} to obtain some analytical insight regarding the effective interaction $V_{\bm{q}}$ and the parameter $\lambda$.
In the limit of a thin ring, we can drop the weak correction at small $q$ and evaluate the logarithm to obtain
\begin{align}
    \Pi_{0,\bm{q}}&=
    \frac{N}{4\pi^2}\frac{k_0^4}{\varepsilon_0}
    \Re\biggl[
    -\frac{i\pi}{q q'}(-i\pi)
    \biggr]
\end{align}
The susceptibility then diverges around $q=0$, so that one can safely take the strong interaction limit, yielding
\begin{align}
    V_{\bm{q}}&=\frac{4}{N}
    \frac{k_0^4}{\varepsilon_0}
    \Re\bigl[qq'\bigr].
\end{align}
For $n\to 0$ (i.e. $\bar\mu\to 0$), the momentum transfer becomes approximately $qq'=k_0^2|\sin\theta|$.
Inserting these simplifications into the linearized gap equation yields for the d-wave channel the matrix elements
\begin{align}\label{eq:pwaveexpansion}
    \mathcal{M}_{11}&=
    \mathcal{M}_{22}=
    \mathcal{M}_{12}=\frac{k_0^2}{3 \pi^3 n}.
\end{align}
The matrix then has the largest eigenvalue $\lambda=2\mathcal{M}_{22}$, which diverges like $n^{-1}$.
In the vicinity of $\bar\mu\to k_0^4$, in the p-wave channel $\lambda$ can also be recovered analytically. We approximate the susceptibility as
\begin{align}
    \Pi_{0,\bm{q}}&=
    \frac{N}{4\pi^2}
    \frac{k_0^4}{\varepsilon_0}
    \Re\biggl[
    \frac{-i\pi}{q q'}\log\frac{q^2-6k_0^2-q'\sqrt{8k_0^2-q^2}}{2k_0^2-q^2+i q q'}
    \biggr].
\end{align}
Inserting the limiting values for $k_{F2}=0$ and $k_{F1}=2k_0$ results in $\Pi^{-1}_{11}=0$ and
\begin{align}
    \frac{N}{4\pi^2}\frac{k_0^4}{\varepsilon_0}
    \Pi^{-1}_{12}&=
    \frac{4k_0^2}{\pi^2}\\
    \frac{N}{4\pi^2}\frac{k_0^4}{\varepsilon_0}
    \Pi^{-1}_{22}&=
    \frac{4k_0^2}{\pi}\Im\biggl[
    \frac{\log(\alpha_+(\theta)\alpha_-(\theta))}
    {\sqrt{2\cos\theta}|\sin\frac{\theta}{2}|}
    \biggr]^{-1},\label{eq:longform}
\end{align}
where $\alpha_\pm(\theta)=(1\pm 2\cos\theta+2\sqrt{\cos\theta\pm 1}\sqrt{\cos\theta})$. The last expression is captured reasonably well by $\sim -4\sqrt{2}k_0^2|\sin\frac{\theta}{2}|/(\pi \log(3-2\sqrt{2}))$, but we will forego this simplification. Integration in the p-wave channel using Eq.~\eqref{eq:longform} results in the matrix elements $\mathcal{M}_{11}=\mathcal{M}_{12}=0$ and $\mathcal{M}_{22}=0.37/N$.
The largest eigenvalue is therefore $\lambda=\mathcal{M}_{22}$, in excellent agreement with a numerical evaluation.

We note that the solution for a radial dispersion with $k_0^2<0$ [Eq.~\eqref{eq:radsymmain2}], which is a radially symmetric dispersion more representative of the electron side, results in really small values for $\lambda<0.01$, far below the ones found using the tight-binding dispersion (Fig.~\ref{fig-le}). We thus conclude that the latter finding is a parametric effect which depends strongly on the amount of trigonal warping in the system and on the proximity to the VHS. 

\section{Selection of chiral superconductivity below $T_c$}

\begin{figure}[b]
    \centering
    \includegraphics[width=0.18\columnwidth]{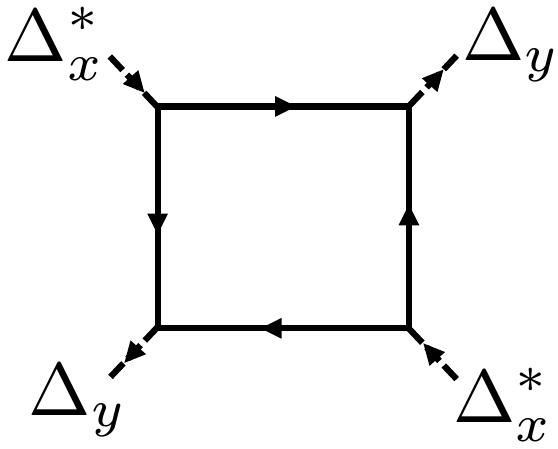}
    \caption{Diagram for the calculation of the quartic term $u_1$ in Eq. (\ref{eq:GL}).}
    \label{fig:quartic}
\end{figure}

In this Section, we discuss the nature of the p--wave superconducting state. At $T_c$, there are two degenerate components of the order parameter, which we denote by $\Delta_x$ and $\Delta_y$. Below $T_c$, these components can either combine into a chiral $\Delta_x + i\Delta_y$ state that breaks time reversal symmetry, or into a nematic superconducting $\alpha \Delta_x + \beta \Delta_y$ state (where $\alpha$ and $\beta$ are real), breaking $C_3$ rotational symmetry. To determine which state is preferred, we need to solve the nonlinear BCS gap equation and compare their condensation energies. Near $T_c$, this can be done by considering the quartic terms in the Landau free energy functional. Up to fourth order in $\Delta_{x,y}$, the most general free energy density compatible with $C_3$ and time reversal symmetries can be written as
\begin{equation}
f = r(T)\left(|\Delta_x|^2 + |\Delta_y|^2\right) + \frac{u}{2}\left(|\Delta_x|^2 + |\Delta_y|^2\right)^2
+\frac{u_1}{2}\left(\Delta^*_x \Delta_y - \Delta^*_y \Delta_x\right)^2
\label{eq:GL}
\end{equation}
where, as usual, $r(T) \propto (T - T_c)/T_c$, and stability requires that $u>0$ $u_1>-u$. $u_1>0$ ($u_1<0$) favors chiral (nematic) SC, respectively. 

Within BCS theory, the quartic terms are determined by the non-interacting band structure. The sign of $u_1$ can be determined by expanding the second term in Eq. (\ref{eq:GL}) and examining the $(\Delta^{*}_{x})^2 \Delta_{y}^2$ contribution. This contribution comes from the diagram shown in Fig.~\ref{fig:quartic}. Denoting the solutions of the linearized gap equation [Eq.~(4) in the main text] as $\Delta_{a,\bm{k}} = \Delta_a f_a(\bm{k})$ with $a=x,y$ and $f_{x,y}(\bm{k})$ are real functions (such a choice is possible due to time reversal symmetry), we can express $u_1$ as 
\begin{equation}
    u_1 = \frac{1}{2} T\sum_{\omega_n}\int\frac{d^2 k}{(2\pi)^2} \frac{f_x^2(\bm{k}) f_y^2(\bm{k})}{(\omega_n^2 + \varepsilon^2_{\bm{k},+})^2}>0.
\end{equation}
Here, $\omega_n = \pi(2n+1)T$ are Matsubara frequencies. Hence, within our model, the chiral SC is favored below $T_c$. Note that this conclusion relies only on the $C_3$ symmetry, and does not depend on any microscopic details of the system.

\section{Role of Hund's coupling}
The Hund's term lifts the degeneracy between the singlet and triplet SC states. We assume the following simple form for $H_{{\rm Hund}}$:
\begin{equation}
    H_{{\rm Hund}}=-\int d^{2}r\int d^{2}r'\,J_{H}(\bm{r}-\bm{r}')\bm{S}_{+}(\bm{r})\cdot\bm{S}_{-}(\bm{r}').
\end{equation}
Here, $\bm{S}_{\tau}(\bm{r})$ is the spin density in valley $\tau$,
\begin{equation}
    S_{\tau}^{j}(\bm{r})=\frac{1}{2}\sum_{\ell,\sigma}\sum_{\alpha,\beta=\uparrow,\downarrow}\psi_{\ell,\sigma,\tau,\alpha}^{\dagger}(\bm{r})\,s_{\alpha\beta}^{j}\,\psi_{\ell,\sigma,\tau,\beta}(\bm{r}),
\end{equation}
where $j=x,y,z$ and $s^j$ are Pauli matrices in spin space. $J_H(\bm{r})$ decays rapidly in real space. Naively, its typical range is the lattice spacing $a$, since microscopically, the Hund's coupling originates from lattice-scale interactions.
As mentioned in the main text, $J_H$ receives contributions from both Coulomb and electron-phonon interactions, and may have either sign.

Projecting $H_{\rm{Hund}}$ to the valence band, it is useful to rewrite it in the Cooper channel as
\begin{equation}
    H_{{\rm Hund}}=-\frac{1}{8L^{2}}\sum_{\bm{k},\bm{k}',\bm{q}}\tilde{J}_{H}(\bm{k}-\bm{k}')\Lambda_{\bm{k}',\bm{k},+1}\Lambda_{-\bm{k}'+\bm{q},-\bm{k}+\bm{q},-1}\left[3\bm{\Phi}_{t}^{\dagger}(\bm{k}',\bm{q})\cdot\bm{\Phi}_{t}(\bm{k},\bm{q})-\Phi_{s}^{\dagger}(\bm{k}',\bm{q})\Phi_{s}(\bm{k},\bm{q})\right],
\end{equation}
where $\tilde{J}_H(\bm{q}) = \int d^2r\,e^{-i\bm{q}\cdot\bm{r}} J_H(\bm{r})$, and 
\begin{align}
\bm{\Phi}_{t}(\bm{k},\bm{q})&=c_{-\bm{k}+\bm{q},-}^{T}\left(is_{y}\bm{s}\right)c_{\bm{k},+},\nonumber\\
{\Phi}_{s}(\bm{k},\bm{q})&=c_{-\bm{k}+\bm{q},-}^{T}\left(is_{y}\right)c_{\bm{k},+}.
\end{align}
Here, $c^T_{\bm{k},\tau} = (c_{\bm{k},\tau,\uparrow}, c_{\bm{k},\tau,\downarrow})$, and $\Phi_s$, $\bm{\Phi}_t$ are the singlet and triplet pairing operators, respectively. 

The Hund's term is incorporated in the linearized gap equation [\eqref{eq:lin}] by adding to the linear operator $\mathcal{M}$ a term $\mathcal{M}_H$ defined by its action on the singlet and triplet gap functions, $(\Delta_{s})_{\bm{k}}$ and $(\bm{\Delta}_{t})_{\bm{k}}$ respectively, as
\begin{align}
    (\mathcal{M}_{\rm{H}} \Delta_{s})_{\bm{k}}&= -\frac{1}{8}\int \frac{dk'_\parallel}{(2\pi)^2 v_{\bm{k}'}} \tilde{J}_H(\bm{k} - \bm{k}') |\Lambda_{\bm{k},\bm{k}-\bm{k}',+1}|^2(\Delta_s)_{\bm{k}'},\nonumber\\
    (\mathcal{M}_{\rm{H}} \bm{\Delta}_{t})_{\bm{k}}&= \frac{3}{8}\int \frac{dk'_\parallel}{(2\pi)^2 v_{\bm{k}'}} \tilde{J}_H(\bm{k} - \bm{k}') |\Lambda_{\bm{k},\bm{k}-\bm{k}',+1}|^2(\bm{\Delta}_t)_{\bm{k}'}.
    \label{eq:MH}
\end{align}
The dimensionless SC coupling constants for singlet and triplet pairing, $\lambda_{s,t}$, satisfy $(\mathcal{M}+\mathcal{M}_H)\Delta_{s,t} = \lambda_{s,t} \Delta_{s,t}$. 

To zeroth order in $\tilde{J}_H$, the singlet and triplet channels are degenerate. Denoting the solution of the linearized gap equation in this case by $\Delta_{\bm{k}}$,  we can estimate the singlet-triplet splitting $\Delta \lambda_{st} \equiv \lambda_s - \lambda_t$ using first-order perturbation theory as
\begin{equation}
    \Delta\lambda_{st}\approx -\frac{1}{2}\frac{\int_{k_{\parallel}}\int_{k'_{\parallel}}\Delta_{\bm{k}}^{*}\tilde{J}_{H}(\bm{k}-\bm{k}')\left|\Lambda_{\bm{k},\bm{k}-\bm{k}',+1}\right|{}^{2}\Delta_{\bm{k}'}}{\int_{k_{\parallel}}\left|\Delta_{\bm{k}}\right|^{2}},
    \label{eq:DL}
\end{equation}
where we introduced the shorthand notation $\int_{k_{\parallel}}\equiv\int\frac{dk_{\parallel}}{(2\pi)^{2}v_{\bm{k}}}$. 
First, let us neglect the momentum dependence of $\tilde{J}_H(\bm{q})$ (i.e., we assume that the Hund's term is local in real space). Then, denoting $\tilde{J}_H(\bm{q}) = J_{H,0}$, we find that $J_{H,0}>0$ always favors triplet pairing ($\Delta \lambda_{st}\le 0$), as expected for a ferromagnetic Hund's coupling.

However, if $\Delta_{\bm{k}}$ has p-wave symmetry, then for a momentum-independent $\tilde{J}_H$ we get $\Delta \lambda_{st}=0$. This is since in this case, $\Delta_{\bm{k}}$ is annihilated by $\mathcal{M}_H$. This can be checked directly using Eq. \eqref{eq:MH}, using the symmetry properties of the system under $C_3$. Lifting the singlet-triplet degeneracy requires $\tilde{J}_H$ to be momentum dependent.
In order to understand how the momentum dependence of $\tilde{J}_{H}$ gives rise to a singlet-triplet splitting in the p-wave case, we expand $\tilde{J}_{H}(\bm{q})$ around $\bm{q}=0$, writing it is as
\begin{equation}
    \tilde{J}_H(\bm{q}) = J_{H,0} + J_{H,2} (qa)^2 + O(q^4). 
\end{equation}
Note that no other terms are allowed by the $C_3$ symmetry to second order in $\bm{q}$. Assuming that $\Delta_{k}$ has angular momentum $m=1$, i.e. $\Delta_{R_3 \bm{k}} = e^{2i\pi/3}\Delta_{\bm{k}}$ where $R_3$ is a rotation matrix by $2\pi/3$, we can rewrite Eq.~(\ref{eq:DL}) as
\begin{equation}
    \Delta\lambda_{st}\approx \frac{J_{H,2}\,a^2}{4}\frac{\int_{k_{\parallel}}\int_{k'_{\parallel}}\left|\Lambda_{\bm{k},\bm{k}-\bm{k}',+1}\right|^{2}\Delta_{\bm{k}}^{*}(k_x+i k_y)(k'_x - i k'_y)\Delta_{\bm{k}'}}{\int_{k_{\parallel}}\left|\Delta_{\bm{k}}\right|^{2}}.
    \label{eq:DLpwave}
\end{equation}
The integral in the numerator of Eq. \eqref{eq:DLpwave} is non-negative and is not required to vanish by symmetry. Therefore, in the p-wave phase, $J_{H,2}>0$ ($J_{H,2}<0$) favors singlet (triplet) pairing, respectively.

\begin{figure}[t]
    \centering
    \includegraphics[width=.49\columnwidth]{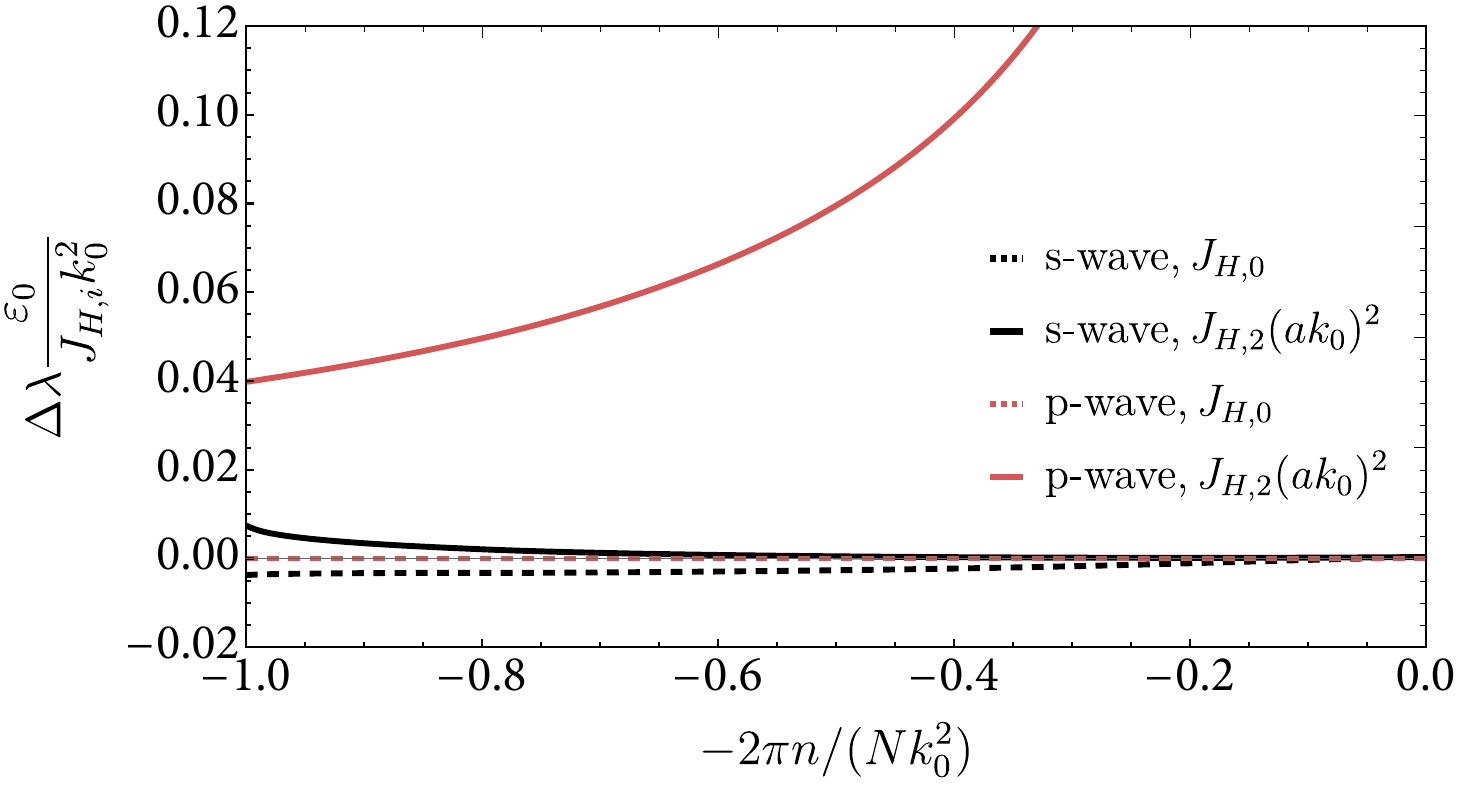}
    \caption{Effect of the Hund's coupling on $\lambda$ for a circularly symmetric dispersion. For small $J_{H}$, the normalized correction $\Delta\lambda$ is proportional to $J_{H}$, namely $\Delta\lambda\propto J_{H,0}$ for s-wave order and $\Delta\lambda\propto J_{H,2}(ak_0)^2$ for p-wave order. As shown in the text, in the case of p-wave, the correction has a divergent coefficient in the limit $n\to 0$. Note that 
    $J_{H,i}$ is normalized in units of density of states $k_0^2/\varepsilon_0$.
    Furthermore, there is a multiplicative factor for the spin configuration depending whether the pairing is in the singlet or triplet channel.}
    \label{fig:MH_dlambda}
\end{figure}

In order to demonstrate the effects of $J_{H,0}$ and $J_{H,2}$ on the singlet-triplet splitting, we show $\Delta \lambda_{st}$ as a function of density in the rotationally-symmetric model for both the p-wave and extended s-wave phases in Fig.~\ref{fig:MH_dlambda}. 
The coefficients of the correction terms are universal in that they do not depend on the flavor number $N$. 
For extended s-wave, the main correction comes from the momentum independent part $J_{H,0}$, which nearly cancels due to the sign change of the order parameter between both Fermi surfaces, leaving only a small magnitude $<0.01 k_0^2J_{H,0}/\varepsilon_0$ (Fig.~\ref{fig:MH_dlambda}). In the case of p-wave order the part from $J_{H,0}$ is identically zero. To determine the next contribution at order $q^2$, we use that for p-wave and for a thin ring it holds approximately that $\mathcal{M}_{11}=\mathcal{M}_{12}=\mathcal{M}_{22}$, making the first perturbation in $J_{H,2}$ independent of the value of $\mathcal{M}$.
The result is then
$\Delta\lambda=\frac{\varepsilon_0}{8\pi \sqrt{\bar\mu}}J_{H,2}(ak_0)^2$,
which diverges upon taking the limit $n\to 0$ ($\sqrt{\bar\mu}\to 0$). Note that the correction to the d-wave channel vanishes identically both for $J_{H,0}$ and $J_{H,2}$. As mentioned before, in the singlet pairing channel the correction due to the Hund's coupling is further multiplied by $-1/4$, while the factor is $+3/4$ for the triplet channel.


\begin{thebibliography}{53}%
\makeatletter
\providecommand \@ifxundefined [1]{%
 \@ifx{#1\undefined}
}%
\providecommand \@ifnum [1]{%
 \ifnum #1\expandafter \@firstoftwo
 \else \expandafter \@secondoftwo
 \fi
}%
\providecommand \@ifx [1]{%
 \ifx #1\expandafter \@firstoftwo
 \else \expandafter \@secondoftwo
 \fi
}%
\providecommand \natexlab [1]{#1}%
\providecommand \enquote  [1]{``#1''}%
\providecommand \bibnamefont  [1]{#1}%
\providecommand \bibfnamefont [1]{#1}%
\providecommand \citenamefont [1]{#1}%
\providecommand \href@noop [0]{\@secondoftwo}%
\providecommand \href [0]{\begingroup \@sanitize@url \@href}%
\providecommand \@href[1]{\@@startlink{#1}\@@href}%
\providecommand \@@href[1]{\endgroup#1\@@endlink}%
\providecommand \@sanitize@url [0]{\catcode `\\12\catcode `\$12\catcode
  `\&12\catcode `\#12\catcode `\^12\catcode `\_12\catcode `\%12\relax}%
\providecommand \@@startlink[1]{}%
\providecommand \@@endlink[0]{}%
\providecommand \url  [0]{\begingroup\@sanitize@url \@url }%
\providecommand \@url [1]{\endgroup\@href {#1}{\urlprefix }}%
\providecommand \urlprefix  [0]{URL }%
\providecommand \Eprint [0]{\href }%
\providecommand \doibase [0]{https://doi.org/}%
\providecommand \selectlanguage [0]{\@gobble}%
\providecommand \bibinfo  [0]{\@secondoftwo}%
\providecommand \bibfield  [0]{\@secondoftwo}%
\providecommand \translation [1]{[#1]}%
\providecommand \BibitemOpen [0]{}%
\providecommand \bibitemStop [0]{}%
\providecommand \bibitemNoStop [0]{.\EOS\space}%
\providecommand \EOS [0]{\spacefactor3000\relax}%
\providecommand \BibitemShut  [1]{\csname bibitem#1\endcsname}%
\let\auto@bib@innerbib\@empty
\bibitem [{\citenamefont {Feldman}\ \emph {et~al.}(2009)\citenamefont
  {Feldman}, \citenamefont {Martin},\ and\ \citenamefont
  {Yacoby}}]{feldman2009broken}%
  \BibitemOpen
  \bibfield  {author} {\bibinfo {author} {\bibfnamefont {B.~E.}\ \bibnamefont
  {Feldman}}, \bibinfo {author} {\bibfnamefont {J.}~\bibnamefont {Martin}},\
  and\ \bibinfo {author} {\bibfnamefont {A.}~\bibnamefont {Yacoby}},\
  }\bibfield  {title} {\bibinfo {title} {Broken-symmetry states and divergent
  resistance in suspended bilayer graphene},\ }\href@noop {} {\bibfield
  {journal} {\bibinfo  {journal} {Nature Physics}\ }\textbf {\bibinfo {volume}
  {5}},\ \bibinfo {pages} {889} (\bibinfo {year} {2009})}\BibitemShut {NoStop}%
\bibitem [{\citenamefont {Mayorov}\ \emph {et~al.}(2011)\citenamefont
  {Mayorov}, \citenamefont {Elias}, \citenamefont {Mucha-Kruczynski},
  \citenamefont {Gorbachev}, \citenamefont {Tudorovskiy}, \citenamefont
  {Zhukov}, \citenamefont {Morozov}, \citenamefont {Katsnelson}, \citenamefont
  {Fal'ko}, \citenamefont {Geim} \emph {et~al.}}]{mayorov2011interaction}%
  \BibitemOpen
  \bibfield  {author} {\bibinfo {author} {\bibfnamefont {A.}~\bibnamefont
  {Mayorov}}, \bibinfo {author} {\bibfnamefont {D.}~\bibnamefont {Elias}},
  \bibinfo {author} {\bibfnamefont {M.}~\bibnamefont {Mucha-Kruczynski}},
  \bibinfo {author} {\bibfnamefont {R.}~\bibnamefont {Gorbachev}}, \bibinfo
  {author} {\bibfnamefont {T.}~\bibnamefont {Tudorovskiy}}, \bibinfo {author}
  {\bibfnamefont {A.}~\bibnamefont {Zhukov}}, \bibinfo {author} {\bibfnamefont
  {S.}~\bibnamefont {Morozov}}, \bibinfo {author} {\bibfnamefont
  {M.}~\bibnamefont {Katsnelson}}, \bibinfo {author} {\bibfnamefont
  {V.}~\bibnamefont {Fal'ko}}, \bibinfo {author} {\bibfnamefont
  {A.}~\bibnamefont {Geim}}, \emph {et~al.},\ }\bibfield  {title} {\bibinfo
  {title} {Interaction-driven spectrum reconstruction in bilayer graphene},\
  }\href@noop {} {\bibfield  {journal} {\bibinfo  {journal} {Science}\ }\textbf
  {\bibinfo {volume} {333}},\ \bibinfo {pages} {860} (\bibinfo {year}
  {2011})}\BibitemShut {NoStop}%
\bibitem [{\citenamefont {Cao}\ \emph {et~al.}(2018{\natexlab{a}})\citenamefont
  {Cao}, \citenamefont {Fatemi}, \citenamefont {Demir}, \citenamefont {Fang},
  \citenamefont {Tomarken}, \citenamefont {Luo}, \citenamefont
  {Sanchez-Yamagishi}, \citenamefont {Watanabe}, \citenamefont {Taniguchi},
  \citenamefont {Kaxiras} \emph {et~al.}}]{cao2018correlated}%
  \BibitemOpen
  \bibfield  {author} {\bibinfo {author} {\bibfnamefont {Y.}~\bibnamefont
  {Cao}}, \bibinfo {author} {\bibfnamefont {V.}~\bibnamefont {Fatemi}},
  \bibinfo {author} {\bibfnamefont {A.}~\bibnamefont {Demir}}, \bibinfo
  {author} {\bibfnamefont {S.}~\bibnamefont {Fang}}, \bibinfo {author}
  {\bibfnamefont {S.~L.}\ \bibnamefont {Tomarken}}, \bibinfo {author}
  {\bibfnamefont {J.~Y.}\ \bibnamefont {Luo}}, \bibinfo {author} {\bibfnamefont
  {J.~D.}\ \bibnamefont {Sanchez-Yamagishi}}, \bibinfo {author} {\bibfnamefont
  {K.}~\bibnamefont {Watanabe}}, \bibinfo {author} {\bibfnamefont
  {T.}~\bibnamefont {Taniguchi}}, \bibinfo {author} {\bibfnamefont
  {E.}~\bibnamefont {Kaxiras}}, \emph {et~al.},\ }\bibfield  {title} {\bibinfo
  {title} {Correlated insulator behaviour at half-filling in magic-angle
  graphene superlattices},\ }\href {https://doi.org/10.1038/nature26154}
  {\bibfield  {journal} {\bibinfo  {journal} {Nature}\ }\textbf {\bibinfo
  {volume} {556}},\ \bibinfo {pages} {80} (\bibinfo {year}
  {2018}{\natexlab{a}})}\BibitemShut {NoStop}%
\bibitem [{\citenamefont {Cao}\ \emph {et~al.}(2018{\natexlab{b}})\citenamefont
  {Cao}, \citenamefont {Fatemi}, \citenamefont {Fang}, \citenamefont
  {Watanabe}, \citenamefont {Taniguchi}, \citenamefont {Kaxiras},\ and\
  \citenamefont {Jarillo-Herrero}}]{cao2018unconventional}%
  \BibitemOpen
  \bibfield  {author} {\bibinfo {author} {\bibfnamefont {Y.}~\bibnamefont
  {Cao}}, \bibinfo {author} {\bibfnamefont {V.}~\bibnamefont {Fatemi}},
  \bibinfo {author} {\bibfnamefont {S.}~\bibnamefont {Fang}}, \bibinfo {author}
  {\bibfnamefont {K.}~\bibnamefont {Watanabe}}, \bibinfo {author}
  {\bibfnamefont {T.}~\bibnamefont {Taniguchi}}, \bibinfo {author}
  {\bibfnamefont {E.}~\bibnamefont {Kaxiras}},\ and\ \bibinfo {author}
  {\bibfnamefont {P.}~\bibnamefont {Jarillo-Herrero}},\ }\bibfield  {title}
  {\bibinfo {title} {Unconventional superconductivity in magic-angle graphene
  superlattices},\ }\href {https://doi.org/10.1038/nature26160} {\bibfield
  {journal} {\bibinfo  {journal} {Nature}\ }\textbf {\bibinfo {volume} {556}},\
  \bibinfo {pages} {43} (\bibinfo {year} {2018}{\natexlab{b}})}\BibitemShut
  {NoStop}%
\bibitem [{\citenamefont {Lu}\ \emph {et~al.}(2019)\citenamefont {Lu},
  \citenamefont {Stepanov}, \citenamefont {Yang}, \citenamefont {Xie},
  \citenamefont {Aamir}, \citenamefont {Das}, \citenamefont {Urgell},
  \citenamefont {Watanabe}, \citenamefont {Taniguchi}, \citenamefont {Zhang}
  \emph {et~al.}}]{lu2019superconductors}%
  \BibitemOpen
  \bibfield  {author} {\bibinfo {author} {\bibfnamefont {X.}~\bibnamefont
  {Lu}}, \bibinfo {author} {\bibfnamefont {P.}~\bibnamefont {Stepanov}},
  \bibinfo {author} {\bibfnamefont {W.}~\bibnamefont {Yang}}, \bibinfo {author}
  {\bibfnamefont {M.}~\bibnamefont {Xie}}, \bibinfo {author} {\bibfnamefont
  {M.~A.}\ \bibnamefont {Aamir}}, \bibinfo {author} {\bibfnamefont
  {I.}~\bibnamefont {Das}}, \bibinfo {author} {\bibfnamefont {C.}~\bibnamefont
  {Urgell}}, \bibinfo {author} {\bibfnamefont {K.}~\bibnamefont {Watanabe}},
  \bibinfo {author} {\bibfnamefont {T.}~\bibnamefont {Taniguchi}}, \bibinfo
  {author} {\bibfnamefont {G.}~\bibnamefont {Zhang}}, \emph {et~al.},\
  }\bibfield  {title} {\bibinfo {title} {Superconductors, orbital magnets and
  correlated states in magic-angle bilayer graphene},\ }\href
  {https://doi.org/10.1038/s41586-019-1695-0} {\bibfield  {journal} {\bibinfo
  {journal} {Nature}\ }\textbf {\bibinfo {volume} {574}},\ \bibinfo {pages}
  {653} (\bibinfo {year} {2019})}\BibitemShut {NoStop}%
\bibitem [{\citenamefont {Liu}\ \emph {et~al.}(2020)\citenamefont {Liu},
  \citenamefont {Hao}, \citenamefont {Khalaf}, \citenamefont {Lee},
  \citenamefont {Ronen}, \citenamefont {Yoo}, \citenamefont {Najafabadi},
  \citenamefont {Watanabe}, \citenamefont {Taniguchi}, \citenamefont
  {Vishwanath} \emph {et~al.}}]{liu2019spin}%
  \BibitemOpen
  \bibfield  {author} {\bibinfo {author} {\bibfnamefont {X.}~\bibnamefont
  {Liu}}, \bibinfo {author} {\bibfnamefont {Z.}~\bibnamefont {Hao}}, \bibinfo
  {author} {\bibfnamefont {E.}~\bibnamefont {Khalaf}}, \bibinfo {author}
  {\bibfnamefont {J.~Y.}\ \bibnamefont {Lee}}, \bibinfo {author} {\bibfnamefont
  {Y.}~\bibnamefont {Ronen}}, \bibinfo {author} {\bibfnamefont
  {H.}~\bibnamefont {Yoo}}, \bibinfo {author} {\bibfnamefont {D.~H.}\
  \bibnamefont {Najafabadi}}, \bibinfo {author} {\bibfnamefont
  {K.}~\bibnamefont {Watanabe}}, \bibinfo {author} {\bibfnamefont
  {T.}~\bibnamefont {Taniguchi}}, \bibinfo {author} {\bibfnamefont
  {A.}~\bibnamefont {Vishwanath}}, \emph {et~al.},\ }\bibfield  {title}
  {\bibinfo {title} {Tunable spin-polarized correlated states in twisted double
  bilayer graphene},\ }\href
  {https://doi.org/https://doi.org/10.1038/s41586-020-2458-7} {\bibfield
  {journal} {\bibinfo  {journal} {Nature}\ }\textbf {\bibinfo {volume} {583}},\
  \bibinfo {pages} {221} (\bibinfo {year} {2020})}\BibitemShut {NoStop}%
\bibitem [{\citenamefont {Yankowitz}\ \emph {et~al.}(2019)\citenamefont
  {Yankowitz}, \citenamefont {Chen}, \citenamefont {Polshyn}, \citenamefont
  {Zhang}, \citenamefont {Watanabe}, \citenamefont {Taniguchi}, \citenamefont
  {Graf}, \citenamefont {Young},\ and\ \citenamefont {Dean}}]{Yankowitz1059}%
  \BibitemOpen
  \bibfield  {author} {\bibinfo {author} {\bibfnamefont {M.}~\bibnamefont
  {Yankowitz}}, \bibinfo {author} {\bibfnamefont {S.}~\bibnamefont {Chen}},
  \bibinfo {author} {\bibfnamefont {H.}~\bibnamefont {Polshyn}}, \bibinfo
  {author} {\bibfnamefont {Y.}~\bibnamefont {Zhang}}, \bibinfo {author}
  {\bibfnamefont {K.}~\bibnamefont {Watanabe}}, \bibinfo {author}
  {\bibfnamefont {T.}~\bibnamefont {Taniguchi}}, \bibinfo {author}
  {\bibfnamefont {D.}~\bibnamefont {Graf}}, \bibinfo {author} {\bibfnamefont
  {A.~F.}\ \bibnamefont {Young}},\ and\ \bibinfo {author} {\bibfnamefont
  {C.~R.}\ \bibnamefont {Dean}},\ }\bibfield  {title} {\bibinfo {title} {Tuning
  superconductivity in twisted bilayer graphene},\ }\href
  {https://doi.org/10.1126/science.aav1910} {\bibfield  {journal} {\bibinfo
  {journal} {Science}\ }\textbf {\bibinfo {volume} {363}},\ \bibinfo {pages}
  {1059} (\bibinfo {year} {2019})}\BibitemShut {NoStop}%
\bibitem [{\citenamefont {Chen}\ \emph {et~al.}(2019)\citenamefont {Chen},
  \citenamefont {Sharpe}, \citenamefont {Gallagher}, \citenamefont {Rosen},
  \citenamefont {Fox}, \citenamefont {Jiang}, \citenamefont {Lyu},
  \citenamefont {Li}, \citenamefont {Watanabe}, \citenamefont {Taniguchi} \emph
  {et~al.}}]{chen2019signatures}%
  \BibitemOpen
  \bibfield  {author} {\bibinfo {author} {\bibfnamefont {G.}~\bibnamefont
  {Chen}}, \bibinfo {author} {\bibfnamefont {A.~L.}\ \bibnamefont {Sharpe}},
  \bibinfo {author} {\bibfnamefont {P.}~\bibnamefont {Gallagher}}, \bibinfo
  {author} {\bibfnamefont {I.~T.}\ \bibnamefont {Rosen}}, \bibinfo {author}
  {\bibfnamefont {E.~J.}\ \bibnamefont {Fox}}, \bibinfo {author} {\bibfnamefont
  {L.}~\bibnamefont {Jiang}}, \bibinfo {author} {\bibfnamefont
  {B.}~\bibnamefont {Lyu}}, \bibinfo {author} {\bibfnamefont {H.}~\bibnamefont
  {Li}}, \bibinfo {author} {\bibfnamefont {K.}~\bibnamefont {Watanabe}},
  \bibinfo {author} {\bibfnamefont {T.}~\bibnamefont {Taniguchi}}, \emph
  {et~al.},\ }\bibfield  {title} {\bibinfo {title} {Signatures of tunable
  superconductivity in a trilayer graphene moir{\'e} superlattice},\ }\href
  {https://www.nature.com/articles/s41586-019-1393-y#citeas} {\bibfield
  {journal} {\bibinfo  {journal} {Nature}\ }\textbf {\bibinfo {volume} {572}},\
  \bibinfo {pages} {215} (\bibinfo {year} {2019})}\BibitemShut {NoStop}%
\bibitem [{\citenamefont {Hao}\ \emph {et~al.}(2021)\citenamefont {Hao},
  \citenamefont {Zimmerman}, \citenamefont {Ledwith}, \citenamefont {Khalaf},
  \citenamefont {Najafabadi}, \citenamefont {Watanabe}, \citenamefont
  {Taniguchi}, \citenamefont {Vishwanath},\ and\ \citenamefont
  {Kim}}]{hao2021electric}%
  \BibitemOpen
  \bibfield  {author} {\bibinfo {author} {\bibfnamefont {Z.}~\bibnamefont
  {Hao}}, \bibinfo {author} {\bibfnamefont {A.}~\bibnamefont {Zimmerman}},
  \bibinfo {author} {\bibfnamefont {P.}~\bibnamefont {Ledwith}}, \bibinfo
  {author} {\bibfnamefont {E.}~\bibnamefont {Khalaf}}, \bibinfo {author}
  {\bibfnamefont {D.~H.}\ \bibnamefont {Najafabadi}}, \bibinfo {author}
  {\bibfnamefont {K.}~\bibnamefont {Watanabe}}, \bibinfo {author}
  {\bibfnamefont {T.}~\bibnamefont {Taniguchi}}, \bibinfo {author}
  {\bibfnamefont {A.}~\bibnamefont {Vishwanath}},\ and\ \bibinfo {author}
  {\bibfnamefont {P.}~\bibnamefont {Kim}},\ }\bibfield  {title} {\bibinfo
  {title} {Electric field--tunable superconductivity in alternating-twist
  magic-angle trilayer graphene},\ }\href@noop {} {\bibfield  {journal}
  {\bibinfo  {journal} {Science}\ }\textbf {\bibinfo {volume} {371}},\ \bibinfo
  {pages} {1133} (\bibinfo {year} {2021})}\BibitemShut {NoStop}%
\bibitem [{\citenamefont {Park}\ \emph {et~al.}(2021)\citenamefont {Park},
  \citenamefont {Cao}, \citenamefont {Watanabe}, \citenamefont {Taniguchi},\
  and\ \citenamefont {Jarillo-Herrero}}]{park2021tunable}%
  \BibitemOpen
  \bibfield  {author} {\bibinfo {author} {\bibfnamefont {J.~M.}\ \bibnamefont
  {Park}}, \bibinfo {author} {\bibfnamefont {Y.}~\bibnamefont {Cao}}, \bibinfo
  {author} {\bibfnamefont {K.}~\bibnamefont {Watanabe}}, \bibinfo {author}
  {\bibfnamefont {T.}~\bibnamefont {Taniguchi}},\ and\ \bibinfo {author}
  {\bibfnamefont {P.}~\bibnamefont {Jarillo-Herrero}},\ }\bibfield  {title}
  {\bibinfo {title} {Tunable strongly coupled superconductivity in magic-angle
  twisted trilayer graphene},\ }\href@noop {} {\bibfield  {journal} {\bibinfo
  {journal} {Nature}\ }\textbf {\bibinfo {volume} {590}},\ \bibinfo {pages}
  {249} (\bibinfo {year} {2021})}\BibitemShut {NoStop}%
\bibitem [{\citenamefont {Koshino}\ and\ \citenamefont
  {McCann}(2009)}]{Koshino2009}%
  \BibitemOpen
  \bibfield  {author} {\bibinfo {author} {\bibfnamefont {M.}~\bibnamefont
  {Koshino}}\ and\ \bibinfo {author} {\bibfnamefont {E.}~\bibnamefont
  {McCann}},\ }\bibfield  {title} {\bibinfo {title} {Trigonal warping and
  berry's phase $n\ensuremath{\pi}$ in {ABC}-stacked multilayer graphene},\
  }\href {https://doi.org/10.1103/PhysRevB.80.165409} {\bibfield  {journal}
  {\bibinfo  {journal} {Phys. Rev. B}\ }\textbf {\bibinfo {volume} {80}},\
  \bibinfo {pages} {165409} (\bibinfo {year} {2009})}\BibitemShut {NoStop}%
\bibitem [{\citenamefont {Zhang}\ \emph {et~al.}(2010)\citenamefont {Zhang},
  \citenamefont {Sahu}, \citenamefont {Min},\ and\ \citenamefont
  {MacDonald}}]{MacDonald2010}%
  \BibitemOpen
  \bibfield  {author} {\bibinfo {author} {\bibfnamefont {F.}~\bibnamefont
  {Zhang}}, \bibinfo {author} {\bibfnamefont {B.}~\bibnamefont {Sahu}},
  \bibinfo {author} {\bibfnamefont {H.}~\bibnamefont {Min}},\ and\ \bibinfo
  {author} {\bibfnamefont {A.~H.}\ \bibnamefont {MacDonald}},\ }\bibfield
  {title} {\bibinfo {title} {Band structure of {ABC}-stacked graphene
  trilayers},\ }\href {https://doi.org/10.1103/PhysRevB.82.035409} {\bibfield
  {journal} {\bibinfo  {journal} {Phys. Rev. B}\ }\textbf {\bibinfo {volume}
  {82}},\ \bibinfo {pages} {035409} (\bibinfo {year} {2010})}\BibitemShut
  {NoStop}%
\bibitem [{\citenamefont {Koshino}(2010)}]{Koshino2010}%
  \BibitemOpen
  \bibfield  {author} {\bibinfo {author} {\bibfnamefont {M.}~\bibnamefont
  {Koshino}},\ }\bibfield  {title} {\bibinfo {title} {Interlayer screening
  effect in graphene multilayers with {ABA} and {ABC} stacking},\ }\href
  {https://doi.org/10.1103/PhysRevB.81.125304} {\bibfield  {journal} {\bibinfo
  {journal} {Phys. Rev. B}\ }\textbf {\bibinfo {volume} {81}},\ \bibinfo
  {pages} {125304} (\bibinfo {year} {2010})}\BibitemShut {NoStop}%
\bibitem [{\citenamefont {Mak}\ \emph {et~al.}(2010)\citenamefont {Mak},
  \citenamefont {Shan},\ and\ \citenamefont {Heinz}}]{Mak2010}%
  \BibitemOpen
  \bibfield  {author} {\bibinfo {author} {\bibfnamefont {K.~F.}\ \bibnamefont
  {Mak}}, \bibinfo {author} {\bibfnamefont {J.}~\bibnamefont {Shan}},\ and\
  \bibinfo {author} {\bibfnamefont {T.~F.}\ \bibnamefont {Heinz}},\ }\bibfield
  {title} {\bibinfo {title} {Electronic structure of few-layer graphene:
  Experimental demonstration of strong dependence on stacking sequence},\
  }\href {https://doi.org/10.1103/PhysRevLett.104.176404} {\bibfield  {journal}
  {\bibinfo  {journal} {Phys. Rev. Lett.}\ }\textbf {\bibinfo {volume} {104}},\
  \bibinfo {pages} {176404} (\bibinfo {year} {2010})}\BibitemShut {NoStop}%
\bibitem [{\citenamefont {Zhou}\ \emph
  {et~al.}(2021{\natexlab{a}})\citenamefont {Zhou}, \citenamefont {Xie},
  \citenamefont {Ghazaryan}, \citenamefont {Holder}, \citenamefont {Ehrets},
  \citenamefont {Spanton}, \citenamefont {Taniguchi}, \citenamefont {Watanabe},
  \citenamefont {Berg}, \citenamefont {Serbyn},\ and\ \citenamefont
  {Young}}]{Zhou2021half}%
  \BibitemOpen
  \bibfield  {author} {\bibinfo {author} {\bibfnamefont {H.}~\bibnamefont
  {Zhou}}, \bibinfo {author} {\bibfnamefont {T.}~\bibnamefont {Xie}}, \bibinfo
  {author} {\bibfnamefont {A.}~\bibnamefont {Ghazaryan}}, \bibinfo {author}
  {\bibfnamefont {T.}~\bibnamefont {Holder}}, \bibinfo {author} {\bibfnamefont
  {J.~R.}\ \bibnamefont {Ehrets}}, \bibinfo {author} {\bibfnamefont {E.~M.}\
  \bibnamefont {Spanton}}, \bibinfo {author} {\bibfnamefont {T.}~\bibnamefont
  {Taniguchi}}, \bibinfo {author} {\bibfnamefont {K.}~\bibnamefont {Watanabe}},
  \bibinfo {author} {\bibfnamefont {E.}~\bibnamefont {Berg}}, \bibinfo {author}
  {\bibfnamefont {M.}~\bibnamefont {Serbyn}}, \ and\ \bibinfo {author}
  {\bibfnamefont {A.~F.}\ \bibnamefont {Young}},\ }\bibfield  {title} {\bibinfo {title} {{Half and quarter metals in rhombohedral trilayer graphene}},\ }\href
  {https://doi.org/10.1038/s41586-021-03938-w} {\bibfield  {journal} {\bibinfo
  {journal} {Nature}\ }\textbf {\bibinfo {volume} {598}},\ \bibinfo {pages}
  {429} (\bibinfo {year} {2021}{\natexlab{a}})}\BibitemShut {NoStop}%
\bibitem [{\citenamefont {Zhou}\ \emph
  {et~al.}(2021{\natexlab{b}})\citenamefont {Zhou}, \citenamefont {Xie},
  \citenamefont {Taniguchi}, \citenamefont {Watanabe},\ and\ \citenamefont
  {Young}}]{Zhou2021superconductivity}%
  \BibitemOpen
  \bibfield  {author} {\bibinfo {author} {\bibfnamefont {H.}~\bibnamefont
  {Zhou}}, \bibinfo {author} {\bibfnamefont {T.}~\bibnamefont {Xie}}, \bibinfo
  {author} {\bibfnamefont {T.}~\bibnamefont {Taniguchi}}, \bibinfo {author}
  {\bibfnamefont {K.}~\bibnamefont {Watanabe}}, \ and\ \bibinfo {author}
  {\bibfnamefont {A.~F.}\ \bibnamefont {Young}},\ }\bibfield  {title} {\bibinfo {title} {{Superconductivity in rhombohedral trilayer graphene}},\ }\href
  {https://doi.org/10.1038/s41586-021-03926-0} {\bibfield  {journal} {\bibinfo
  {journal} {Nature}\ }\textbf {\bibinfo {volume} {598}},\ \bibinfo {pages}
  {434} (\bibinfo {year} {2021}{\natexlab{b}})}\BibitemShut {NoStop}%
\bibitem [{not()}]{note-unconventional}%
  \BibitemOpen
  \href@noop {} {}\bibinfo {note} {For a related mechanism for unconventional
  superconductivity in trilayer graphene, starting from the Hubbard model, see
  H. Dai, J. Hou, X. Zhang, Y. Liang, and T. Ma, Phys. Rev. B \textbf{104},
  035104 (2021).}\BibitemShut {Stop}%
\bibitem [{\citenamefont {Kohn}\ and\ \citenamefont
  {Luttinger}(1965)}]{KohnLuttinger}%
  \BibitemOpen
  \bibfield  {author} {\bibinfo {author} {\bibfnamefont {W.}~\bibnamefont
  {Kohn}}\ and\ \bibinfo {author} {\bibfnamefont {J.~M.}\ \bibnamefont
  {Luttinger}},\ }\bibfield  {title} {\bibinfo {title} {New mechanism for
  superconductivity},\ }\href {https://doi.org/10.1103/PhysRevLett.15.524}
  {\bibfield  {journal} {\bibinfo  {journal} {Phys. Rev. Lett.}\ }\textbf
  {\bibinfo {volume} {15}},\ \bibinfo {pages} {524} (\bibinfo {year}
  {1965})}\BibitemShut {NoStop}%
\bibitem [{\citenamefont {Kagan}\ \emph {et~al.}(2015)\citenamefont {Kagan},
  \citenamefont {Mitskan},\ and\ \citenamefont
  {Korovushkin}}]{kagan2015anomalous}%
  \BibitemOpen
  \bibfield  {author} {\bibinfo {author} {\bibfnamefont {M.~Y.}\ \bibnamefont
  {Kagan}}, \bibinfo {author} {\bibfnamefont {V.~A.}\ \bibnamefont {Mitskan}},\
  and\ \bibinfo {author} {\bibfnamefont {M.~M.}\ \bibnamefont {Korovushkin}},\
  }\bibfield  {title} {\bibinfo {title} {Anomalous superconductivity and
  superfluidity in repulsive fermion systems},\ }\href
  {https://iopscience.iop.org/article/10.3367/UFNe.0185.201508a.0785}
  {\bibfield  {journal} {\bibinfo  {journal} {Physics-Uspekhi}\ }\textbf
  {\bibinfo {volume} {58}},\ \bibinfo {pages} {733} (\bibinfo {year}
  {2015})}\BibitemShut {NoStop}%
\bibitem [{\citenamefont {Raghu}\ \emph {et~al.}(2010)\citenamefont {Raghu},
  \citenamefont {Kivelson},\ and\ \citenamefont {Scalapino}}]{Raghu2010}%
  \BibitemOpen
  \bibfield  {author} {\bibinfo {author} {\bibfnamefont {S.}~\bibnamefont
  {Raghu}}, \bibinfo {author} {\bibfnamefont {S.~A.}\ \bibnamefont
  {Kivelson}},\ and\ \bibinfo {author} {\bibfnamefont {D.~J.}\ \bibnamefont
  {Scalapino}},\ }\bibfield  {title} {\bibinfo {title} {{Superconductivity in
  the repulsive Hubbard model: An asymptotically exact weak-coupling
  solution}},\ }\href {https://doi.org/10.1103/PhysRevB.81.224505} {\bibfield
  {journal} {\bibinfo  {journal} {Phys. Rev. B}\ }\textbf {\bibinfo {volume}
  {81}},\ \bibinfo {pages} {224505} (\bibinfo {year} {2010})}\BibitemShut
  {NoStop}%
\bibitem [{\citenamefont {Maiti}\ and\ \citenamefont
  {Chubukov}(2013)}]{maiti2013superconductivity}%
  \BibitemOpen
  \bibfield  {author} {\bibinfo {author} {\bibfnamefont {S.}~\bibnamefont
  {Maiti}}\ and\ \bibinfo {author} {\bibfnamefont {A.~V.}\ \bibnamefont
  {Chubukov}},\ }\bibfield  {title} {\bibinfo {title} {Superconductivity from
  repulsive interaction},\ }in\ \href
  {https://aip.scitation.org/doi/abs/10.1063/1.4818400} {\emph {\bibinfo
  {booktitle} {AIP Conference Proceedings}}},\ Vol.\ \bibinfo {volume} {1550}\
  (\bibinfo {organization} {American Institute of Physics},\ \bibinfo {year}
  {2013})\ pp.\ \bibinfo {pages} {3--73}\BibitemShut {NoStop}%
\bibitem [{com()}]{comment_two_band}%
  \BibitemOpen
  \href@noop {} {}\bibinfo {note} {Similarly, a two-dimensional band structure
  with multiple sub-bands has been predicted to be favorable for an
  unconventional mechanism for superconductivity
  \cite{Raghu2011,Chubukov2017}.}\BibitemShut {Stop}%
\bibitem [{\citenamefont {Raghu}\ and\ \citenamefont
  {Kivelson}(2011)}]{Raghu2011}%
  \BibitemOpen
  \bibfield  {author} {\bibinfo {author} {\bibfnamefont {S.}~\bibnamefont
  {Raghu}}\ and\ \bibinfo {author} {\bibfnamefont {S.~A.}\ \bibnamefont
  {Kivelson}},\ }\bibfield  {title} {\bibinfo {title} {Superconductivity from
  repulsive interactions in the two-dimensional electron gas},\ }\href
  {https://doi.org/10.1103/PhysRevB.83.094518} {\bibfield  {journal} {\bibinfo
  {journal} {Phys. Rev. B}\ }\textbf {\bibinfo {volume} {83}},\ \bibinfo
  {pages} {094518} (\bibinfo {year} {2011})}\BibitemShut {NoStop}%
\bibitem [{\citenamefont {Chubukov}\ and\ \citenamefont
  {Kivelson}(2017)}]{Chubukov2017}%
  \BibitemOpen
  \bibfield  {author} {\bibinfo {author} {\bibfnamefont {A.~V.}\ \bibnamefont
  {Chubukov}}\ and\ \bibinfo {author} {\bibfnamefont {S.~A.}\ \bibnamefont
  {Kivelson}},\ }\bibfield  {title} {\bibinfo {title} {Superconductivity in
  engineered two-dimensional electron gases},\ }\href
  {https://doi.org/10.1103/PhysRevB.96.174514} {\bibfield  {journal} {\bibinfo
  {journal} {Phys. Rev. B}\ }\textbf {\bibinfo {volume} {96}},\ \bibinfo
  {pages} {174514} (\bibinfo {year} {2017})}\BibitemShut {NoStop}%
\bibitem [{\citenamefont {Mazin}\ \emph {et~al.}(2008)\citenamefont {Mazin},
  \citenamefont {Singh}, \citenamefont {Johannes},\ and\ \citenamefont
  {Du}}]{Mazin2008}%
  \BibitemOpen
  \bibfield  {author} {\bibinfo {author} {\bibfnamefont {I.~I.}\ \bibnamefont
  {Mazin}}, \bibinfo {author} {\bibfnamefont {D.~J.}\ \bibnamefont {Singh}},
  \bibinfo {author} {\bibfnamefont {M.~D.}\ \bibnamefont {Johannes}},\ and\
  \bibinfo {author} {\bibfnamefont {M.~H.}\ \bibnamefont {Du}},\ }\bibfield
  {title} {\bibinfo {title} {{Unconventional Superconductivity with a Sign
  Reversal in the Order Parameter of
  ${\mathrm{LaFeAsO}}_{1\ensuremath{-}x}{\mathrm{F}}_{x}$}},\ }\href
  {https://doi.org/10.1103/PhysRevLett.101.057003} {\bibfield  {journal}
  {\bibinfo  {journal} {Phys. Rev. Lett.}\ }\textbf {\bibinfo {volume} {101}},\
  \bibinfo {pages} {057003} (\bibinfo {year} {2008})}\BibitemShut {NoStop}%
\bibitem [{\citenamefont {{Zondiner}}\ \emph {et~al.}(2020)\citenamefont
  {{Zondiner}}, \citenamefont {{Rozen}}, \citenamefont {{Rodan-Legrain}},
  \citenamefont {{Cao}}, \citenamefont {{Queiroz}}, \citenamefont
  {{Taniguchi}}, \citenamefont {{Watanabe}}, \citenamefont {{Oreg}},
  \citenamefont {{von Oppen}}, \citenamefont {{Stern}}, \citenamefont {{Berg}},
  \citenamefont {{Jarillo-Herrero}},\ and\ \citenamefont
  {{Ilani}}}]{Zondiner2019}%
  \BibitemOpen
  \bibfield  {author} {\bibinfo {author} {\bibfnamefont {U.}~\bibnamefont
  {{Zondiner}}}, \bibinfo {author} {\bibfnamefont {A.}~\bibnamefont {{Rozen}}},
  \bibinfo {author} {\bibfnamefont {D.}~\bibnamefont {{Rodan-Legrain}}},
  \bibinfo {author} {\bibfnamefont {Y.}~\bibnamefont {{Cao}}}, \bibinfo
  {author} {\bibfnamefont {R.}~\bibnamefont {{Queiroz}}}, \bibinfo {author}
  {\bibfnamefont {T.}~\bibnamefont {{Taniguchi}}}, \bibinfo {author}
  {\bibfnamefont {K.}~\bibnamefont {{Watanabe}}}, \bibinfo {author}
  {\bibfnamefont {Y.}~\bibnamefont {{Oreg}}}, \bibinfo {author} {\bibfnamefont
  {F.}~\bibnamefont {{von Oppen}}}, \bibinfo {author} {\bibfnamefont
  {A.}~\bibnamefont {{Stern}}}, \bibinfo {author} {\bibfnamefont
  {E.}~\bibnamefont {{Berg}}}, \bibinfo {author} {\bibfnamefont
  {P.}~\bibnamefont {{Jarillo-Herrero}}},\ and\ \bibinfo {author}
  {\bibfnamefont {S.}~\bibnamefont {{Ilani}}},\ }\bibfield  {title} {\bibinfo
  {title} {{Cascade of phase transitions and Dirac revivals in magic-angle
  graphene}},\ }\href {https://doi.org/10.1038/s41586-020-2373-y} {\bibfield
  {journal} {\bibinfo  {journal} {\nat}\ }\textbf {\bibinfo {volume} {582}},\
  \bibinfo {pages} {203} (\bibinfo {year} {2020})}\BibitemShut {NoStop}%
\bibitem [{\citenamefont {Wong}\ \emph {et~al.}(2020)\citenamefont {Wong},
  \citenamefont {Nuckolls}, \citenamefont {Oh}, \citenamefont {Lian},
  \citenamefont {Xie}, \citenamefont {Jeon}, \citenamefont {Watanabe},
  \citenamefont {Taniguchi}, \citenamefont {Bernevig},\ and\ \citenamefont
  {Yazdani}}]{Wong2020cascade}%
  \BibitemOpen
  \bibfield  {author} {\bibinfo {author} {\bibfnamefont {D.}~\bibnamefont
  {Wong}}, \bibinfo {author} {\bibfnamefont {K.~P.}\ \bibnamefont {Nuckolls}},
  \bibinfo {author} {\bibfnamefont {M.}~\bibnamefont {Oh}}, \bibinfo {author}
  {\bibfnamefont {B.}~\bibnamefont {Lian}}, \bibinfo {author} {\bibfnamefont
  {Y.}~\bibnamefont {Xie}}, \bibinfo {author} {\bibfnamefont {S.}~\bibnamefont
  {Jeon}}, \bibinfo {author} {\bibfnamefont {K.}~\bibnamefont {Watanabe}},
  \bibinfo {author} {\bibfnamefont {T.}~\bibnamefont {Taniguchi}}, \bibinfo
  {author} {\bibfnamefont {B.~A.}\ \bibnamefont {Bernevig}},\ and\ \bibinfo
  {author} {\bibfnamefont {A.}~\bibnamefont {Yazdani}},\ }\bibfield  {title}
  {\bibinfo {title} {Cascade of electronic transitions in magic-angle twisted
  bilayer graphene},\ }\href
  {https://doi.org/https://doi.org/10.1038/s41586-020-2339-0} {\bibfield
  {journal} {\bibinfo  {journal} {Nature}\ }\textbf {\bibinfo {volume} {582}},\
  \bibinfo {pages} {198} (\bibinfo {year} {2020})}\BibitemShut {NoStop}%
\bibitem [{\citenamefont {Clogston}(1962)}]{Clogston}%
  \BibitemOpen
  \bibfield  {author} {\bibinfo {author} {\bibfnamefont {A.~M.}\ \bibnamefont
  {Clogston}},\ }\bibfield  {title} {\bibinfo {title} {Upper limit for the
  critical field in hard superconductors},\ }\href
  {https://doi.org/10.1103/PhysRevLett.9.266} {\bibfield  {journal} {\bibinfo
  {journal} {Phys. Rev. Lett.}\ }\textbf {\bibinfo {volume} {9}},\ \bibinfo
  {pages} {266} (\bibinfo {year} {1962})}\BibitemShut {NoStop}%
\bibitem [{\citenamefont {Chandrasekhar}(1962)}]{chandrasekhar1962note}%
  \BibitemOpen
  \bibfield  {author} {\bibinfo {author} {\bibfnamefont {B.~S.}\ \bibnamefont
  {Chandrasekhar}},\ }\bibfield  {title} {\bibinfo {title} {A note on the
  maximum critical field of high-field superconductors},\ }\href@noop {}
  {\bibfield  {journal} {\bibinfo  {journal} {Applied Physics Letters}\
  }\textbf {\bibinfo {volume} {1}},\ \bibinfo {pages} {7} (\bibinfo {year}
  {1962})}\BibitemShut {NoStop}%
\bibitem [{\citenamefont {Chou}\ \emph {et~al.}(2021)\citenamefont {Chou},
  \citenamefont {Wu}, \citenamefont {Sau},\ and\ \citenamefont
  {Sarma}}]{Chou2021}%
  \BibitemOpen
  \bibfield  {author} {\bibinfo {author} {\bibfnamefont {Y.-Z.}\ \bibnamefont
  {Chou}}, \bibinfo {author} {\bibfnamefont {F.}~\bibnamefont {Wu}}, \bibinfo
  {author} {\bibfnamefont {J.~D.}\ \bibnamefont {Sau}}, \ and\ \bibinfo
  {author} {\bibfnamefont {S.~D.}\ \bibnamefont {Sarma}},\ }\bibfield  {title} {\bibinfo {title} {{Acoustic-phonon-mediated
  superconductivity in rhombohedral trilayer graphene}},\ }\href
  {https://doi.org/10.1103/PhysRevLett.127.187001} {\bibfield  {journal}
  {\bibinfo  {journal} {Phys. Rev. Lett.}\ }\textbf {\bibinfo {volume}
  {127}},\ \bibinfo {pages} {187001} (\bibinfo {year} {2021})}\BibitemShut
  {NoStop}%
\bibitem [{\citenamefont {Allen}(2000)}]{allen2000electron}%
  \BibitemOpen
  \bibfield  {author} {\bibinfo {author} {\bibfnamefont {P.~B.}\ \bibnamefont
  {Allen}},\ }\bibfield  {title} {\bibinfo {title} {The electron-phonon
  coupling constant},\ }\href
  {http://felix.physics.sunysb.edu/~allen/Pdffiles/poole.pdf} {\bibfield
  {journal} {\bibinfo  {journal} {Handbook of Superconductivity, edited by
  Charles P. Poole Jr. (Academic Press, San Diego)}\ }\textbf {\bibinfo
  {volume} {9G}},\ \bibinfo {pages} {478} (\bibinfo {year} {2000})}\BibitemShut
  {NoStop}%
\bibitem [{Note1()}]{Note1}%
  \BibitemOpen
  \bibinfo {note} {A possible resolution is that that $J_{H}$ changes its sign
  as a function of density. However, since $J_{H}$ is a lattice-scale
  interaction, it seems unlikely to changes significantly upon a small change
  in electron concentration.}\BibitemShut {Stop}%
\bibitem [{SI()}]{SI}%
  \BibitemOpen
  \href@noop {} {}\bibinfo {note} {See Supplemental Material for more details
  on the band structure and polarization function of RTG and the circularly
  symmetric model, the method for calculation of $T_c$, the SC state below
  $T_c$, and the Hund's coupling.}\BibitemShut {Stop}%
\bibitem [{\citenamefont {Chubukov}(1993)}]{Chubukov1993}%
  \BibitemOpen
  \bibfield  {author} {\bibinfo {author} {\bibfnamefont {A.~V.}\ \bibnamefont
  {Chubukov}},\ }\bibfield  {title} {\bibinfo {title} {{Kohn-Luttinger effect
  and the instability of a two-dimensional repulsive Fermi liquid at $T=0$}},\
  }\href {https://doi.org/10.1103/PhysRevB.48.1097} {\bibfield  {journal}
  {\bibinfo  {journal} {Phys. Rev. B}\ }\textbf {\bibinfo {volume} {48}},\
  \bibinfo {pages} {1097} (\bibinfo {year} {1993})}\BibitemShut {NoStop}%
\bibitem [{\citenamefont {Lee}\ \emph {et~al.}(2019)\citenamefont {Lee},
  \citenamefont {Khalaf}, \citenamefont {Liu}, \citenamefont {Liu},
  \citenamefont {Hao}, \citenamefont {Kim},\ and\ \citenamefont
  {Vishwanath}}]{lee2019theory}%
  \BibitemOpen
  \bibfield  {author} {\bibinfo {author} {\bibfnamefont {J.~Y.}\ \bibnamefont
  {Lee}}, \bibinfo {author} {\bibfnamefont {E.}~\bibnamefont {Khalaf}},
  \bibinfo {author} {\bibfnamefont {S.}~\bibnamefont {Liu}}, \bibinfo {author}
  {\bibfnamefont {X.}~\bibnamefont {Liu}}, \bibinfo {author} {\bibfnamefont
  {Z.}~\bibnamefont {Hao}}, \bibinfo {author} {\bibfnamefont {P.}~\bibnamefont
  {Kim}},\ and\ \bibinfo {author} {\bibfnamefont {A.}~\bibnamefont
  {Vishwanath}},\ }\bibfield  {title} {\bibinfo {title} {Theory of correlated
  insulating behaviour and spin-triplet superconductivity in twisted double
  bilayer graphene},\ }\href {https://doi.org/10.1038/s41467-019-12981-1}
  {\bibfield  {journal} {\bibinfo  {journal} {Nature communications}\ }\textbf
  {\bibinfo {volume} {10}},\ \bibinfo {pages} {1} (\bibinfo {year}
  {2019})}\BibitemShut {NoStop}%
\bibitem [{\citenamefont {{Khalaf}}\ \emph {et~al.}(2020)\citenamefont
  {{Khalaf}}, \citenamefont {{Ledwith}},\ and\ \citenamefont
  {{Vishwanath}}}]{Khalaf2020}%
  \BibitemOpen
  \bibfield  {author} {\bibinfo {author} {\bibfnamefont {E.}~\bibnamefont
  {{Khalaf}}}, \bibinfo {author} {\bibfnamefont {P.}~\bibnamefont
  {{Ledwith}}},\ and\ \bibinfo {author} {\bibfnamefont {A.}~\bibnamefont
  {{Vishwanath}}},\ }\bibfield  {title} {\bibinfo {title} {{Symmetry
  constraints on superconductivity in twisted bilayer graphene: Fractional
  vortices, $4e$ condensates or non-unitary pairing}},\ }\href@noop {}
  {\bibfield  {journal} {\bibinfo  {journal} {arXiv e-prints}\ ,\ \bibinfo
  {eid} {arXiv:2012.05915}} (\bibinfo {year} {2020})},\ \Eprint
  {https://arxiv.org/abs/2012.05915} {arXiv:2012.05915 [cond-mat.supr-con]}
  \BibitemShut {NoStop}%
\bibitem [{\citenamefont {{Yuan}}\ \emph {et~al.}(2019)\citenamefont {{Yuan}},
  \citenamefont {{Isobe}},\ and\ \citenamefont {{Fu}}}]{Yuan2019}%
  \BibitemOpen
  \bibfield  {author} {\bibinfo {author} {\bibfnamefont {N.~F.~Q.}\
  \bibnamefont {{Yuan}}}, \bibinfo {author} {\bibfnamefont {H.}~\bibnamefont
  {{Isobe}}},\ and\ \bibinfo {author} {\bibfnamefont {L.}~\bibnamefont
  {{Fu}}},\ }\bibfield  {title} {\bibinfo {title} {{Magic of high-order van
  Hove singularity}},\ }\href {https://doi.org/10.1038/s41467-019-13670-9}
  {\bibfield  {journal} {\bibinfo  {journal} {Nat. Commun.}\ }\textbf {\bibinfo
  {volume} {10}},\ \bibinfo {eid} {5769} (\bibinfo {year} {2019})},\ \Eprint
  {https://arxiv.org/abs/1901.05432} {arXiv:1901.05432 [cond-mat.str-el]}
  \BibitemShut {NoStop}%
\bibitem [{com()}]{comment_valley}%
  \BibitemOpen
  \href@noop {} {}\bibinfo {note} {We note that due to the trigonal warping in RTG, 
in a each valley, $\varepsilon_{\bm{k},\tau}\neq \varepsilon_{-\bm{k},\tau}$, and hence a valley-polarized state does not have a BCS instability.}\BibitemShut {Stop}%
\bibitem [{Note2()}]{Note2}%
  \BibitemOpen
  \bibinfo {note} {In the case where the FS is composed of disjoint pockets,
  $2k_F$ should replaced by the maximum distance between points on the
  FS.}\BibitemShut {Stop}%
\bibitem [{\citenamefont {Alicea}\ and\ \citenamefont
  {Fisher}(2006)}]{Alicea2006}%
  \BibitemOpen
  \bibfield  {author} {\bibinfo {author} {\bibfnamefont {J.}~\bibnamefont
  {Alicea}}\ and\ \bibinfo {author} {\bibfnamefont {M.~P.~A.}\ \bibnamefont
  {Fisher}},\ }\bibfield  {title} {\bibinfo {title} {Graphene integer quantum
  {Hall} effect in the ferromagnetic and paramagnetic regimes},\ }\href
  {https://doi.org/10.1103/PhysRevB.74.075422} {\bibfield  {journal} {\bibinfo
  {journal} {Phys. Rev. B}\ }\textbf {\bibinfo {volume} {74}},\ \bibinfo
  {pages} {075422} (\bibinfo {year} {2006})}\BibitemShut {NoStop}%
\bibitem [{\citenamefont {Kharitonov}(2012)}]{Kharitonov2012}%
  \BibitemOpen
  \bibfield  {author} {\bibinfo {author} {\bibfnamefont {M.}~\bibnamefont
  {Kharitonov}},\ }\bibfield  {title} {\bibinfo {title} {Phase diagram for the
  $\ensuremath{\nu}=0$ quantum {Hall} state in monolayer graphene},\ }\href
  {https://doi.org/10.1103/PhysRevB.85.155439} {\bibfield  {journal} {\bibinfo
  {journal} {Phys. Rev. B}\ }\textbf {\bibinfo {volume} {85}},\ \bibinfo
  {pages} {155439} (\bibinfo {year} {2012})}\BibitemShut {NoStop}%
\bibitem [{\citenamefont {Anderson}(1959)}]{Anderson1959theory}%
\BibitemOpen
\bibfield  {author} {\bibinfo {author} {\bibfnamefont {P.~W.}\ \bibnamefont
		{Anderson}},\ }\bibfield  {title} {\bibinfo {title} {Theory of dirty
		superconductors},\ }\href@noop {} {\bibfield  {journal} {\bibinfo  {journal}
		{Journal of Physics and Chemistry of Solids}\ }\textbf {\bibinfo {volume}
		{11}},\ \bibinfo {pages} {26} (\bibinfo {year} {1959})}\BibitemShut {NoStop}%
\bibitem [{\citenamefont {Finkel'stein}(1994)}]{Finkelstein1994suppression}%
\BibitemOpen
\bibfield  {author} {\bibinfo {author} {\bibfnamefont {A.~M.}\ \bibnamefont
		{Finkel'stein}},\ }\bibfield  {title} {\bibinfo {title} {Suppression of
		superconductivity in homogeneously disordered systems},\ }\href@noop {}
{\bibfield  {journal} {\bibinfo  {journal} {Physica B: Condensed Matter}\
	}\textbf {\bibinfo {volume} {197}},\ \bibinfo {pages} {636} (\bibinfo {year}
	{1994})}\BibitemShut {NoStop}%
\bibitem [{Abr()}]{Abrikosov}%
  \BibitemOpen
  \href@noop {} {}\bibinfo {note} {A. A. Abrikosov and L. P. Gor'kov,
  Contribution to the theory of superconducting alloys with paramagnetic
  impurities, Zh. Eksp. Teor. Fiz. 39, 1781 (1960) [Sov. Phys. JETP 12, 1243
  (1961).}\BibitemShut {Stop}%
\bibitem [{Lar()}]{Larkin}%
  \BibitemOpen
  \href@noop {} {}\bibinfo {note} {P. I. Larkin, Vector pairing in
  superconductors of small dimensions, Zh. Eksp. Teor. Fiz. Pis'ma Red. 2, 205
  (1965) [Sov. Phys. JETP Lett. 2, 130 (1965)].}\BibitemShut {Stop}%
\bibitem [{\citenamefont {Millis}\ \emph {et~al.}(1988)\citenamefont {Millis},
  \citenamefont {Sachdev},\ and\ \citenamefont {Varma}}]{Millis1988}%
  \BibitemOpen
  \bibfield  {author} {\bibinfo {author} {\bibfnamefont {A.~J.}\ \bibnamefont
  {Millis}}, \bibinfo {author} {\bibfnamefont {S.}~\bibnamefont {Sachdev}},\
  and\ \bibinfo {author} {\bibfnamefont {C.~M.}\ \bibnamefont {Varma}},\
  }\bibfield  {title} {\bibinfo {title} {Inelastic scattering and pair breaking
  in anisotropic and isotropic superconductors},\ }\href
  {https://doi.org/10.1103/PhysRevB.37.4975} {\bibfield  {journal} {\bibinfo
  {journal} {Phys. Rev. B}\ }\textbf {\bibinfo {volume} {37}},\ \bibinfo
  {pages} {4975} (\bibinfo {year} {1988})}\BibitemShut {NoStop}%
\bibitem [{\citenamefont {Radtke}\ \emph {et~al.}(1993)\citenamefont {Radtke},
  \citenamefont {Levin}, \citenamefont {Sch\"uttler},\ and\ \citenamefont
  {Norman}}]{Radtke1993}%
  \BibitemOpen
  \bibfield  {author} {\bibinfo {author} {\bibfnamefont {R.~J.}\ \bibnamefont
  {Radtke}}, \bibinfo {author} {\bibfnamefont {K.}~\bibnamefont {Levin}},
  \bibinfo {author} {\bibfnamefont {H.-B.}\ \bibnamefont {Sch\"uttler}},\ and\
  \bibinfo {author} {\bibfnamefont {M.~R.}\ \bibnamefont {Norman}},\ }\bibfield
   {title} {\bibinfo {title} {{Predictions for impurity-induced
  ${\mathit{T}}_{\mathit{c}}$ suppression in the high-temperature
  superconductors}},\ }\href {https://doi.org/10.1103/PhysRevB.48.653}
  {\bibfield  {journal} {\bibinfo  {journal} {Phys. Rev. B}\ }\textbf {\bibinfo
  {volume} {48}},\ \bibinfo {pages} {653} (\bibinfo {year} {1993})}\BibitemShut
  {NoStop}%
\bibitem [{\citenamefont {Mackenzie}\ \emph {et~al.}(1998)\citenamefont
  {Mackenzie}, \citenamefont {Haselwimmer}, \citenamefont {Tyler},
  \citenamefont {Lonzarich}, \citenamefont {Mori}, \citenamefont {Nishizaki},\
  and\ \citenamefont {Maeno}}]{Mackenzie1998}%
  \BibitemOpen
  \bibfield  {author} {\bibinfo {author} {\bibfnamefont {A.~P.}\ \bibnamefont
  {Mackenzie}}, \bibinfo {author} {\bibfnamefont {R.~K.~W.}\ \bibnamefont
  {Haselwimmer}}, \bibinfo {author} {\bibfnamefont {A.~W.}\ \bibnamefont
  {Tyler}}, \bibinfo {author} {\bibfnamefont {G.~G.}\ \bibnamefont
  {Lonzarich}}, \bibinfo {author} {\bibfnamefont {Y.}~\bibnamefont {Mori}},
  \bibinfo {author} {\bibfnamefont {S.}~\bibnamefont {Nishizaki}},\ and\
  \bibinfo {author} {\bibfnamefont {Y.}~\bibnamefont {Maeno}},\ }\bibfield
  {title} {\bibinfo {title} {{Extremely Strong Dependence of Superconductivity
  on Disorder in ${\mathrm{Sr}}_{2}{\mathrm{RuO}}_{4}$}},\ }\href
  {https://doi.org/10.1103/PhysRevLett.80.161} {\bibfield  {journal} {\bibinfo
  {journal} {Phys. Rev. Lett.}\ }\textbf {\bibinfo {volume} {80}},\ \bibinfo
  {pages} {161} (\bibinfo {year} {1998})}\BibitemShut {NoStop}%
\bibitem [{\citenamefont {Cornfeld}\ \emph {et~al.}(2021)\citenamefont
  {Cornfeld}, \citenamefont {Rudner},\ and\ \citenamefont
  {Berg}}]{Cornfeld2021}%
  \BibitemOpen
  \bibfield  {author} {\bibinfo {author} {\bibfnamefont {E.}~\bibnamefont
  {Cornfeld}}, \bibinfo {author} {\bibfnamefont {M.~S.}\ \bibnamefont
  {Rudner}},\ and\ \bibinfo {author} {\bibfnamefont {E.}~\bibnamefont {Berg}},\
  }\bibfield  {title} {\bibinfo {title} {{Spin-polarized superconductivity:
  Order parameter topology, current dissipation, and multiple-period Josephson
  effect}},\ }\href {https://doi.org/10.1103/PhysRevResearch.3.013051}
  {\bibfield  {journal} {\bibinfo  {journal} {Phys. Rev. Research}\ }\textbf
  {\bibinfo {volume} {3}},\ \bibinfo {pages} {013051} (\bibinfo {year}
  {2021})}\BibitemShut {NoStop}%
\bibitem [{\citenamefont {Furusaki}\ \emph {et~al.}(2001)\citenamefont
  {Furusaki}, \citenamefont {Matsumoto},\ and\ \citenamefont
  {Sigrist}}]{Furusaki2001}%
  \BibitemOpen
  \bibfield  {author} {\bibinfo {author} {\bibfnamefont {A.}~\bibnamefont
  {Furusaki}}, \bibinfo {author} {\bibfnamefont {M.}~\bibnamefont
  {Matsumoto}},\ and\ \bibinfo {author} {\bibfnamefont {M.}~\bibnamefont
  {Sigrist}},\ }\bibfield  {title} {\bibinfo {title} {Spontaneous hall effect
  in a chiral p-wave superconductor},\ }\href
  {https://doi.org/10.1103/PhysRevB.64.054514} {\bibfield  {journal} {\bibinfo
  {journal} {Phys. Rev. B}\ }\textbf {\bibinfo {volume} {64}},\ \bibinfo
  {pages} {054514} (\bibinfo {year} {2001})}\BibitemShut {NoStop}%
\bibitem [{\citenamefont {Stone}\ and\ \citenamefont {Roy}(2004)}]{Stone2004}%
  \BibitemOpen
  \bibfield  {author} {\bibinfo {author} {\bibfnamefont {M.}~\bibnamefont
  {Stone}}\ and\ \bibinfo {author} {\bibfnamefont {R.}~\bibnamefont {Roy}},\
  }\bibfield  {title} {\bibinfo {title} {Edge modes, edge currents, and gauge
  invariance in ${p}_{x}{+ip}_{y}$ superfluids and superconductors},\ }\href
  {https://doi.org/10.1103/PhysRevB.69.184511} {\bibfield  {journal} {\bibinfo
  {journal} {Phys. Rev. B}\ }\textbf {\bibinfo {volume} {69}},\ \bibinfo
  {pages} {184511} (\bibinfo {year} {2004})}\BibitemShut {NoStop}%
\bibitem [{\citenamefont {Sauls}(2011)}]{Sauls2011}%
  \BibitemOpen
  \bibfield  {author} {\bibinfo {author} {\bibfnamefont {J.~A.}\ \bibnamefont
  {Sauls}},\ }\bibfield  {title} {\bibinfo {title} {Surface states, edge
  currents, and the angular momentum of chiral $p$-wave superfluids},\ }\href
  {https://doi.org/10.1103/PhysRevB.84.214509} {\bibfield  {journal} {\bibinfo
  {journal} {Phys. Rev. B}\ }\textbf {\bibinfo {volume} {84}},\ \bibinfo
  {pages} {214509} (\bibinfo {year} {2011})}\BibitemShut {NoStop}%
\bibitem [{\citenamefont {Black-Schaffer}\ and\ \citenamefont
  {Honerkamp}(2014)}]{Black2014chiral}%
  \BibitemOpen
  \bibfield  {author} {\bibinfo {author} {\bibfnamefont {Annica~M}\
  \bibnamefont {Black-Schaffer}}\ and\ \bibinfo {author} {\bibfnamefont
  {Carsten}\ \bibnamefont {Honerkamp}},\ }\bibfield  {title} {\enquote
  {\bibinfo {title} {Chiral d-wave superconductivity in doped graphene},}\
  }\href {\doibase 10.1088/0953-8984/26/42/423201} {\bibfield  {journal}
  {\bibinfo  {journal} {Journal of Physics: Condensed Matter}\ }\textbf
  {\bibinfo {volume} {26}},\ \bibinfo {pages} {423201} (\bibinfo {year}
  {2014})}\BibitemShut {NoStop}%
\bibitem [{\citenamefont {Black-Schaffer}\ and\ \citenamefont
  {Doniach}(2007)}]{Black2007resonating}%
  \BibitemOpen
  \bibfield  {author} {\bibinfo {author} {\bibfnamefont {Annica~M}\
  \bibnamefont {Black-Schaffer}}\ and\ \bibinfo {author} {\bibfnamefont
  {Sebastian}\ \bibnamefont {Doniach}},\ }\bibfield  {title} {\enquote
  {\bibinfo {title} {Resonating valence bonds and mean-field d-wave
  superconductivity in graphite},}\ }\href {\doibase
  10.1103/PhysRevB.75.134512} {\bibfield  {journal} {\bibinfo  {journal}
  {Physical Review B}\ }\textbf {\bibinfo {volume} {75}},\ \bibinfo {pages}
  {134512} (\bibinfo {year} {2007})}\BibitemShut {NoStop}%
\bibitem [{\citenamefont {Wu}\ \emph {et~al.}()\citenamefont {Wu},
  \citenamefont {Scherer}, \citenamefont {Honerkamp},\ and\ \citenamefont
  {Le~Hur}}]{Wu2013correlated}%
  \BibitemOpen
  \bibfield  {author} {\bibinfo {author} {\bibfnamefont {Wei}\ \bibnamefont
  {Wu}}, \bibinfo {author} {\bibfnamefont {Michael~M}\ \bibnamefont {Scherer}},
  \bibinfo {author} {\bibfnamefont {Carsten}\ \bibnamefont {Honerkamp}}, \ and\
  \bibinfo {author} {\bibfnamefont {Karyn}\ \bibnamefont {Le~Hur}},\ }\bibfield
   {title} {\enquote {\bibinfo {title} {Correlated dirac particles and
  superconductivity on the honeycomb lattice},}\ }\href {\doibase
  10.1103/PhysRevB.87.094521} {\bibfield  {journal} {\bibinfo  {journal}
  {Physical Review B}\ }\textbf {\bibinfo {volume} {87}},\ \bibinfo {pages}
  {094521}}\BibitemShut {NoStop}%
\bibitem [{\citenamefont {Nandkishore}\ \emph {et~al.}()\citenamefont
  {Nandkishore}, \citenamefont {Levitov},\ and\ \citenamefont
  {Chubukov}}]{Nandkishore2012chiral}%
  \BibitemOpen
  \bibfield  {author} {\bibinfo {author} {\bibfnamefont {Rahul}\ \bibnamefont
  {Nandkishore}}, \bibinfo {author} {\bibfnamefont {LS}~\bibnamefont
  {Levitov}}, \ and\ \bibinfo {author} {\bibfnamefont {AV}~\bibnamefont
  {Chubukov}},\ }\bibfield  {title} {\enquote {\bibinfo {title} {Chiral
  superconductivity from repulsive interactions in doped graphene},}\ }\href
  {\doibase 10.1038/nphys2208} {\bibfield  {journal} {\bibinfo  {journal}
  {Nature Physics}\ }\textbf {\bibinfo {volume} {8}},\ \bibinfo {pages}
  {158--163}}\BibitemShut {NoStop}%
\bibitem [{\citenamefont {Wang}\ \emph {et~al.}(2012)\citenamefont {Wang},
  \citenamefont {Xiang}, \citenamefont {Wang}, \citenamefont {Wang},
  \citenamefont {Yang},\ and\ \citenamefont {Lee}}]{Wang2012functional}%
  \BibitemOpen
  \bibfield  {author} {\bibinfo {author} {\bibfnamefont {Wan-Sheng}\
  \bibnamefont {Wang}}, \bibinfo {author} {\bibfnamefont {Yuan-Yuan}\
  \bibnamefont {Xiang}}, \bibinfo {author} {\bibfnamefont {Qiang-Hua}\
  \bibnamefont {Wang}}, \bibinfo {author} {\bibfnamefont {Fa}~\bibnamefont
  {Wang}}, \bibinfo {author} {\bibfnamefont {Fan}\ \bibnamefont {Yang}}, \ and\
  \bibinfo {author} {\bibfnamefont {Dung-Hai}\ \bibnamefont {Lee}},\ }\bibfield
   {title} {\enquote {\bibinfo {title} {Functional renormalization group and
  variational monte carlo studies of the electronic instabilities in graphene
  near 1 4 doping},}\ }\href {\doibase 10.1103/PhysRevB.85.035414} {\bibfield
  {journal} {\bibinfo  {journal} {Physical Review B}\ }\textbf {\bibinfo
  {volume} {85}},\ \bibinfo {pages} {035414} (\bibinfo {year}
  {2012})}\BibitemShut {NoStop}%
\bibitem [{\citenamefont {Kiesel}\ \emph {et~al.}(2012)\citenamefont {Kiesel},
  \citenamefont {Platt}, \citenamefont {Hanke}, \citenamefont {Abanin},\ and\
  \citenamefont {Thomale}}]{Kiesel2012competing}%
  \BibitemOpen
  \bibfield  {author} {\bibinfo {author} {\bibfnamefont {Maximilian~L}\
  \bibnamefont {Kiesel}}, \bibinfo {author} {\bibfnamefont {Christian}\
  \bibnamefont {Platt}}, \bibinfo {author} {\bibfnamefont {Werner}\
  \bibnamefont {Hanke}}, \bibinfo {author} {\bibfnamefont {Dmitry~A}\
  \bibnamefont {Abanin}}, \ and\ \bibinfo {author} {\bibfnamefont {Ronny}\
  \bibnamefont {Thomale}},\ }\bibfield  {title} {\enquote {\bibinfo {title}
  {Competing many-body instabilities and unconventional superconductivity in
  graphene},}\ }\href {\doibase 10.1103/PhysRevB.86.020507} {\bibfield
  {journal} {\bibinfo  {journal} {Physical Review B}\ }\textbf {\bibinfo
  {volume} {86}},\ \bibinfo {pages} {020507} (\bibinfo {year}
  {2012})}
  \BibitemShut {NoStop}%
\bibitem [{\citenamefont {{Chatterjee}}\ \emph {et~al.}(2021)\citenamefont
	{{Chatterjee}}, \citenamefont {{Wang}}, \citenamefont {{Berg}},\ and\
	\citenamefont {{Zaletel}}}]{Chatterjee}%
\BibitemOpen
\bibfield  {author} {\bibinfo {author} {\bibfnamefont {S.}~\bibnamefont
		{{Chatterjee}}}, \bibinfo {author} {\bibfnamefont {T.}~\bibnamefont
		{{Wang}}}, \bibinfo {author} {\bibfnamefont {E.}~\bibnamefont {{Berg}}},\
	and\ \bibinfo {author} {\bibfnamefont {M.~P.}\ \bibnamefont {{Zaletel}}},\
}\bibfield  {title} {\bibinfo {title} {{Inter-valley coherent order and
			isospin fluctuation mediated superconductivity in rhombohedral trilayer
			graphene}},\ }\href@noop {} {\bibfield  {journal} {\bibinfo  {journal} {arXiv
			e-prints}\ ,\ \bibinfo {eid} {arXiv:2109.00002}} (\bibinfo {year} {2021})},\
\Eprint {https://arxiv.org/abs/2109.00002} {arXiv:2109.00002
	[cond-mat.supr-con]} \BibitemShut {NoStop}%
\bibitem [{\citenamefont {{Dong}}\ and\ \citenamefont
	{{Levitov}}(2021)}]{Levitov}%
\BibitemOpen
\bibfield  {author} {\bibinfo {author} {\bibfnamefont {Z.}~\bibnamefont
		{{Dong}}}\ and\ \bibinfo {author} {\bibfnamefont {L.}~\bibnamefont
		{{Levitov}}},\ }\bibfield  {title} {\bibinfo {title} {{Superconductivity in
			the vicinity of an isospin-polarized state in a cubic Dirac band}},\
}\href@noop {} {\bibfield  {journal} {\bibinfo  {journal} {arXiv e-prints}\
		,\ \bibinfo {eid} {arXiv:2109.01133}} (\bibinfo {year} {2021})},\ \Eprint
{https://arxiv.org/abs/2109.01133} {arXiv:2109.01133 [cond-mat.supr-con]}
\end{thebibliography}
\end{document}